\documentclass[letterpaper,twocolumn,10pt]{article}
\usepackage{usenix-2020-09}

\usepackage{tikz}
\usepackage{amsmath}

\usepackage{filecontents}
\usepackage{microtype}
\usepackage[affil-it]{authblk}
\usepackage{graphicx}
\usepackage{hyperref}
\hypersetup{hidelinks}
\usepackage[linesnumbered,lined,vlined,ruled,commentsnumbered,noend]{algorithm2e}
\usepackage{textcomp}
\usepackage{xcolor}
\usepackage{listings}
\usepackage{url}
\usepackage{xspace}
\usepackage{multirow}
\usepackage{balance}
\usepackage{ctable}
\usepackage{listings}
\usepackage{color}
\usepackage{graphicx}
\usepackage{subcaption}
\usepackage[nobiblatex]{xurl}
\usepackage{multirow}
\usepackage{amsmath}
\usepackage{algpseudocode}
\usepackage{caption}
\usepackage{subcaption}

\usepackage{cleveref}
\usepackage{acronym}
\usepackage{fancybox}
\usepackage{verbatim}
\usepackage[normalem]{ulem}
\usepackage{float}
\usepackage{newfloat}
\usepackage{mdframed}
\usepackage{enumitem}
\usepackage{booktabs}
\usepackage{pifont}
\usepackage{tcolorbox}
\usepackage{adjustbox}
\usepackage{amsfonts}
\usepackage{makecell}

\definecolor{red}{RGB}{255,0,0}
\definecolor{pkured}{RGB}{192,0,0}
\definecolor{black}{RGB}{0,0,0}

\newcommand{\eg}{e.g.,\xspace}

\newcommand{\eat}[1]{}

\newcommand{\xmark}{\ding{55}}%
\newcommand{\cmark}{\ding{51}}%
\definecolor{myblue}{RGB}{43, 115, 219}
\newcommand{\bluec}{{\color{myblue}\cmark}}
\newcommand{\redx}{{\color{black}\xmark}}

\newcommand{\revised}[1]{\textcolor{black}{#1}}

\acrodef{gdpr}[GDPR]{General Data Protection Regulation}
\acrodef{pipl}[PIPL]{Personal Information Protection Law}

\newcommand{\alipay}{AliPay\xspace}

\newcommand{\mini}{Mini-H\xspace}
\newcommand{\op}{Op-H\xspace}

\newcommand{\tool}{THEFT\xspace}

\begin{document}

\date{}

\title{I Can Tell Your Secrets: Inferring Privacy Attributes from Mini-app Interaction History in Super-apps}

\author[1]{Yifeng Cai}
\author[2]{Ziqi Zhang}
\author[1]{Mengyu Yao}
\author[1]{Junlin Liu}
\author[3]{Xiaoke Zhao}
\author[3]{Xinyi Fu}
\author[3]{Ruoyu Li}
\author[3]{Zhe Li}
\author[1]{\\Xiangqun Chen}
\author[1]{Yao Guo}
\author[1]{Ding Li}

\affil[1]{MOE Key Lab of HCST (PKU), School of Computer Science, Peking University}
\affil[2]{Department of Computer Science, University of Illinois Urbana-Champaign}
\affil[3]{Ant Group}

\maketitle

\begin{abstract}
Super-apps have emerged as comprehensive platforms integrating various mini-apps to provide diverse services. While super-apps offer convenience and enriched functionality, they can introduce new privacy risks. This paper reveals a new privacy leakage source in super-apps: mini-app interaction history, including mini-app usage history (Mini-H) and operation history (Op-H). Mini-H refers to the history of mini-apps accessed by users, such as their frequency and categories. Op-H captures user interactions within mini-apps, including button clicks, bar drags, and image views.  Super-apps can naturally collect these data without instrumentation due to the web-based feature of mini-apps. We identify these data types as novel and unexplored privacy risks through a literature review of 30 papers and an empirical analysis of 31 super-apps. We design a mini-app interaction history-oriented inference attack (THEFT), to exploit this new vulnerability. Using THEFT, the insider threats within the low-privilege business department of the super-app vendor acting as the adversary can achieve more than 95.5\% accuracy in inferring privacy attributes of over 16.1\% of users. THEFT only requires a small training dataset of 200 users from public breached databases on the Internet. We also engage with super-app vendors and a standards association to increase industry awareness and commitment to protect this data. Our contributions are significant in identifying overlooked privacy risks, demonstrating the effectiveness of a new attack, and influencing industry practices toward better privacy protection in the super-app ecosystem.
\end{abstract}

\vspace{-1em}
\section{Introduction}

\begin{figure}[!t]
    \centering
    \includegraphics[width=0.95\columnwidth]{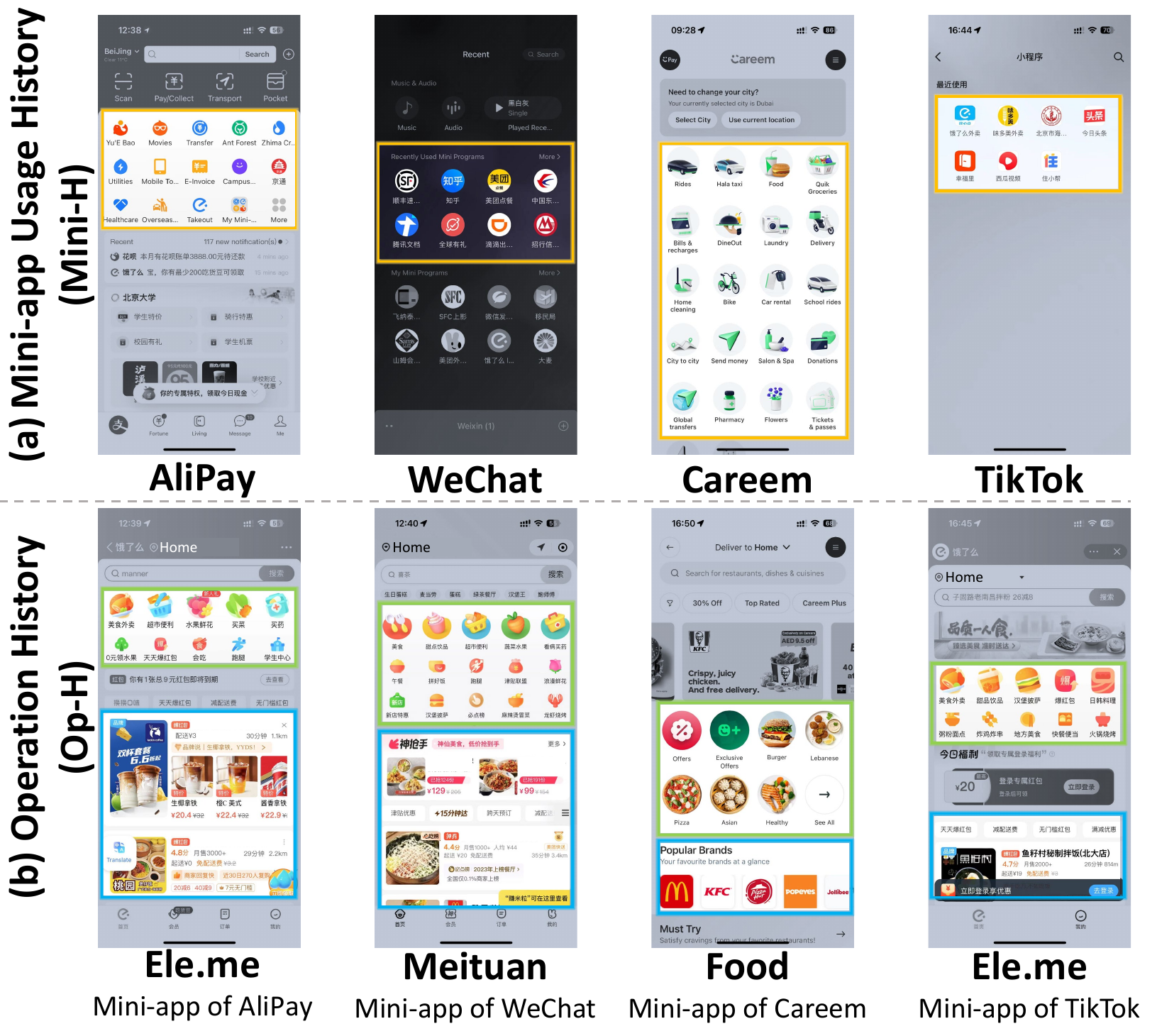}
	\caption{Samples of mini-app interaction history.}
	\label{fig:history}
 \vspace{-1em}
\end{figure}

\textbf{Super-apps.}
With the advances in mobile computing, apps are becoming more and more powerful. Super-apps, such as AliPay~\cite{alipay}, WeChat~\cite{wechat}, and Careem~\cite{careem}, are comprehensive and one-stop applications that allow users to access various mini-apps easily.
Mini-apps are similar to native apps, enabling super-apps to establish an
ecosystem like Google Play and the Apple App Store. This design enriches super-apps'
functionalities and offers great convenience to mobile users. For
example, the food delivery service provider Ele.me~\cite{Eleme} has greatly
benefited from this paradigm: merchants can use mini-apps to sell foods
through AliPay and WeChat directly. Users can browse products, share
interests on social networks, and make purchases without leaving the
super-app. Therefore, super-apps have significantly facilitated daily
activities (e.g., transactions and transportation) from the
built-in or third-party mini-apps. As shown in Figure~\ref{fig:history}~(a), on the homepages of super-apps, the
mini-apps are crucial components and can be easily accessed (highlighted in
the yellow box in Figure~\ref{fig:history}~(a)). 

\noindent\textbf{Blur Definition in Regulations.}
Similar to normal mobile apps, the data collection and storage of super-app is regulated by laws, such as the \ac{gdpr} and the \ac{pipl} \cite{koch2023ok,bielova2024effect,lu2021whois,kollnig2023before}. The laws require mobile apps to provide users (data subjects) with transparent information on the collection and processing of personal data \cite{alizadeh2020gdpr}. The regulations also require super-app vendors to store the data safely and prevent them from transferring the collected data to other parties without the consent of the users~\cite{du2023withdrawing}. However, laws are general. In practice, \textit{what kind of data should be protected depends on the academic community's and industry's research and standard agreement}.

\noindent\textbf{New Leakage Source.}
This paper focuses on an underexplored aspect of data privacy in super-apps: \textit{the mini-app interaction history}. The interactions include various user activities, including clicking buttons, and recently-accessed and preferred mini-apps. A department in the super-app vendor is motivated to use such data to infer user privacy to improve their service~\cite{wu2022linkteller}.
These interaction data are dangerous because \textit{the super-app vendor can naturally collect such data from the released version of super-app due to the design philosophy of mini-apps.} The vendors do not need to modify the app (e.g., instrumentation) to obtain the interaction history data. This is because existing mini-app services are provided online and based on cloud servers according to the development guidance of popular super-apps~\cite{alipaysdk,wechatsdk,tiktoksdk}. 
Specifically, we study two typical types of mini-app interaction history: \textbf{the Mini-app Usage History (\mini)} and \textbf{the Operation History (\op)}. 
\textbf{\mini} includes the mini-apps that the user engaged with over a period of time, which offer clues about user preferences because the frequently accessed mini-app categories can reveal privacy attributes. 
When a user opens a super-app, the super-app displays the frequently accessed mini-apps on the main screen to provide a better user experience. The list of frequently used mini-apps is obtained by sending an HTTP request to the super-app vendor's server, so it is naturally obtained by the super-app provider.
\textbf{\op} includes user interactions within the mini-app, such as button clicks, bar drags, and image views. \op can expose privacy attributes as well. For example, the interaction speed can indicate the user's age information: younger users can navigate and switch interfaces swiftly, whereas older users exhibit slower operation speeds and more repetitive clicks to locate desired information.
Because all mini-apps are web-based applications, once the user clicks a button, super-apps need to send the name and ID of the clicked button to the server to request the mini-app's subsequent feedback~\cite{10.1145/3607199.3607236}. Thus, the super-app vendor can naturally access \op from the request logs on the server.

\noindent\revised{
\noindent\textbf{People's Unawareness.}
We first study how much attention people (academic researchers and industrial practitioners) pay to this new privacy leakage. To perform a comprehensive study, we surveyed 30 papers published in top-tier conferences from 2019 and 31 real-world super-apps. We found that, to concretize laws like \ac{gdpr} and \ac{pipl}, the community has converged on an extensive list of concrete data items considered sensitive. These include personal information and attributes (e.g., name, gender, age) and device information (e.g., IMEI, MAC address). The large number of protected data types shows the efforts of the community to preserve user privacy. 
However, we observed that \textit{neither academia nor industry has fully recognized
the seriousness of the leakage from mini-app interaction history}, let alone the laws and regulations. The literature review shows that none of the existing academic papers has discussed the privacy issues of the mini-app interaction history. The empirical study of super-apps reveals a parallel lack of awareness in the industry. Despite the widespread use of these apps, our analysis indicates that only one app recognized the potential privacy risks associated with \mini. This oversight is particularly concerning because \mini and \op can leak many types of user privacy attributes according to our study.
}

\noindent\revised{
\noindent \textbf{New Attack.}
To show the real threat to user privacy from the mini-app interaction history data, we designed a novel attack called THEFT, \textit{mini-app interac\underline{T}ion \underline{H}istory-ori\underline{E}nted in\underline{F}erence a\underline{T}tack}. 
THEFT leverages Deep Neural Networks (DNNs) to infer users' privacy attributes from their mini-app interaction history. Unlike prior work~\cite{tu2018your} that relies on comprehensive OS-level data (e.g., cellular tower IDs and all HTTP requests), THEFT only needs mini-app interaction history, which is naturally collected by super-apps through their cloud-based services and deemed non-sensitive. This low requirement significantly lowers the bar for insider threats~\cite{6957307} as the adversaries to obtain minimal labeled data (e.g., from privacy breaches~\cite{liu2015cloudy,mayer2021now,thomas2017data}), match mobile phone numbers, and train DNN models. Once trained, these models can infer sensitive privacy attributes of users at scale, posing severe threats to user privacy.
}

\noindent\revised{
\noindent \textbf{Results.}
We conduct a large-scale experiment on the internal data of \alipay, a real-world super-app with more than 1.3 billion users.
Experimental results show that THEFT can achieve more than 95.5\% accuracy in inferring privacy attributes of over 16.1\% of more than 219K users with the training data from only 200 users, making THEFT a practical attack. Our attacks in \alipay itself mean the possible privacy leakage for more than 200 million users. This highlights the substantial risk of mini-app interaction history data. The finding demonstrates the practical implications of our threat model: current privacy measurements in super-apps are insufficient against advanced inference attacks.
}

\begin{table*}[]
\caption{Focused topic and concerned privacy of the papers published in top-tier conferences from 2019 to 2024.}
\label{tab:paper_summary}
\begin{adjustbox}{max width=1.\linewidth}
\begin{tabular}{llll}
\hline
\textbf{Paper} & \textbf{Conference} & \textbf{Focused Topic} & \textbf{Concerned Privacy} \\ \hline
Dong et al.~\cite{dongexploring} & Sec 24 & SDK collected hardware identifiers & Pers./Devi.Info. \\
Pan et al.~\cite{pantrap} & Sec 24 & Online automated privacy policy generators & Contacts, Location, Pers/Devi.Info., Microphone, Camera, Sensors \\
Khandelwal et al.~\cite{khandelwal2023unpacking} & Sec 24 & Data safety sections & Location, Pers./Devi.Info. \\
Klein et al.~\cite{klein2023general} & CCS 23 & Unlawful data processing in web apps & Contacts, Pers.Info. \\
Xiang et al.~\cite{10.1145/3576915.3623067} & CCS 23 & Completeness of privacy policy & Location, Pers.Info. \\
Wang et al.~\cite{10.1145/3576915.3616676} & CCS 23 & Hidden APIs in super-apps & Contacts, Location, Pers./Devi.Info. \\
Zhang et al.~\cite{10.1145/3576915.3616591} & CCS 23 & Leaked master key of mini-apps & Pers.Info. \\
Ferreira et al.~\cite{ferreira2023rulekeeper} & SP 23 & Pers.Info. data compliance in web apps & Location, Pers/Devi.Info. \\
Zhang et al.~\cite{zhang2023understanding} & SP 23 & Face verification system in apps & Camera \\
Xiao et al.~\cite{xiao2023lalaine} & Sec 23 & Compliance of Apple privacy labels & Location, Pers./Devi.Info. \\
Nan et al.~\cite{nan2023are} & Sec 23 & IoT collected data & Devi.Info., Sensors \\
Wang et al.~\cite{10.5555/3620237.3620608} & Sec 23 & APIs in cross platform super-apps & Location, Devi.Info., Microphone, Camera \\
Koch et al.~\cite{koch2023ok} & Sec 23 & Privacy consent dialogs & Location, Devi.Info. \\
Lyons et al.~\cite{lyons2023log} & Sec 23 & Logged personal data in Android & Location, Pers./Devi.Info. \\
Meng et al.~\cite{meng2023post} & NDSS 23 & User-unresettable identifiers in Android & Devi.Info. \\
Jordan et al.~\cite{jordan2021viceroy} & NDSS 23 & Verifiable accountless consumer requests & Contacts, Location, Pers./Devi.Info., Microphone, Camera, Sensors \\
Nguyen et al.~\cite{nguyen2022freely} & CCS 22 & Notice of third-Party tracking in apps & Pers./Devi.Info. \\
Li et al.~\cite{li2022collect} & CCS 22 & Cross-user personal data over-delivery in apps & Pers.Info. \\
Wang et al.~\cite{wang2022srr} & CCS 22 & Location-based service in apps & Location \\
Yang et al.~\cite{yang2022cross} & CCS 22 & Cross mini-app request forgery & Location, Pers/Devi.Info., Microphone, Camera \\
Young et al.~\cite{young2022skilldetective} & Sec 22 & Policy violation of voice assistant apps & Microphone, Camera \\
Balash et al.~\cite{balash2022security} & Sec 22 & Third-party app access for Google account & Pers.Info. \\
Zhang et al.~\cite{zhang2022identity} & Sec 22 & APIs identity confusion in mini-apps & Pers.Info., Contacts \\
Diamantaris et al.~\cite{diamantaris2021sneaky} & CCS 21 & Misuse sensors in apps & Location, Pers.Info., Microphone, Camera, Sensors \\
Bui et al.~\cite{bui2021consistency} & CCS 21 & Consistency of data-usage purposes in apps & Location, Pers./Devi.Info. \\
Haney et al.~\cite{haney2021s} & Sec 21 & Privacy implications & Microphone, Camera \\
Nguyen et al.~\cite{nguyen2021share} & Sec 21 & Personal data compliance in apps & Location, Pers./Devi.Info. \\
\revised{Lu et al.~\cite{lu2020demystifying}} & \revised{CCS 20} & \revised{Security risks of API flaws in mini-apps} & \revised{Location, Pers/Devi.Info., Microphone, Camera} \\
Zuo et al.~\cite{zuo2019does} & SP 19 & Cloud APIs in apps & Pers.Info. \\
Chen et al.~\cite{chen2019demystifying} & SP 19 & Hidden privacy settings in apps & Pers.Info. \\ \hline
\end{tabular}
\end{adjustbox}
\vspace{-0em}
\end{table*}

\noindent \textbf{Generalization.}
Although we mainly use \alipay as a representative super-app to study the
privacy leakage of \mini and \op, our findings are generalizable to
other super-apps. This is because super-apps share
the same \textit{interface design and code structure} across different platforms. 
All mini-apps use JavaScript and WebView to easily replicate in different
super-apps~\cite{li2023minitracker,wang2023taintmini,zhang2022identity}.
Therefore, for different super-apps, mini-apps of different functionalities
often have similar user interfaces and features. Figure \ref{fig:history} (b)
displays four different food delivery mini-apps from four different super-apps.
Green boxes highlight different subcategories that the mini-app can
provide, while blue boxes show the list of stores. The content in both the green
and blue boxes shares similar designs across super-apps. The server's interaction history data collected from these mini-apps are also similar. Therefore,
our attack and findings could apply to any super-apps.

\noindent \textbf{Industrial Feedback.}
We notified 31 super-app vendors and one standards association about the privacy risks of mini-app interaction history. Four vendors agreed to modify the privacy statement of their apps, and the standards association committed to strengthening data protection. Furthermore, these vendors and the association also provided insightful feedback. This feedback ranged from acknowledgments of previously overlooked privacy risks to commitments to enhance data protection measurements. Specifically, developers intended to revise privacy policies and terms on potential dangers and recognize \mini and \op as private data. These responses highlight a growing awareness and proactive stance towards user data privacy in the industry. Overall, the feedback from the industry shows that the mini-app interaction history-oriented inference attack is practical in the real world.

We summarize our contributions as follows.

$\bullet$ We identify the unprotected user mini-app interaction history as a novel underexplored privacy risk in super-apps. We systematically reveal new privacy threats that have been overlooked in academia (through 30 top-tier papers) and industrial practice (through 31 real-world super-apps). 

\vspace{-0.2em}
$\bullet$ We design a mini-app interaction history-oriented inference attack (THEFT), to exploit this privacy risk and demonstrate that the attack can effectively and accurately infer the users' privacy attributes.

\vspace{-0.2em}
$\bullet$ We provide our findings to the super-app vendors to bring their attention to this new privacy risk. We receive valuable feedback acknowledging our contribution to enhancing privacy measurements and the super-app ecosystem.

\vspace{-0.2em}

\section{A Blind Spot in Mini-App Ecosystem}
\label{sec:privacy_of_history}

\revised{In this section, we present the motivation of this paper: people's unawareness of the security risk of mini-app interaction history. Contrary to the vast usage of mini-apps, we found that very few people/companies pay attention to the security risk. To comprehensively evaluate people's consciousness, we perform a thorough literature review in the academic community and an empirical study of industrial practice. Our goal is to present the attitude of both researchers and practitioners. }

\vspace{-1em}
\subsection{Perspective from Academia Community}
\label{sec:academia_community}

\textbf{Methodology.}
To cover the mainstream opinions of the academic community, we survey the papers published in top-tier security conferences \revised{which are included in CSRankings}, including USENIX Security (Sec), IEEE S\&P (SP), ACM CCS (CCS), and NDSS. 
We survey the papers published from 2019 to 2024 \revised{because 2019 is the emergence of super-apps~\cite{zuo2019does}.}
Our literature review consists of three steps. \revised{First, we reviewed the 4,020 accepted papers from the four conferences and selected 43 papers whose titles
relate to the privacy of mobile apps or devices. Then, we read the selected paper's
abstract and introduction sections to filter out 13 papers that did not focus on privacy protection. Finally, we read the full text of the 30 selected papers to summarize the specific types of privacy data that the paper studied. To ensure reliability, four authors (two academic and two industry researchers) independently participated in the above processes, achieving a 99.52\% agreement rate, which indicates a high inter-rater reliability (IRR).}

\begin{table*}[!t]
\centering
\caption{The summary of the collected data types claimed in the privacy policies and terms of 31 super-apps.}
\label{tab:superapp_summary}
\begin{adjustbox}{max width=1.\linewidth}
\begin{tabular}{@{}cccccccccccccc@{}}
\hline
\textbf{No.} & \textbf{Super-app} & \textbf{Location} & \textbf{Contacts} & \textbf{Camera} & \textbf{Gallery} & \textbf{Microphone} & \textbf{Calendar} & \textbf{Devi.} & \textbf{Pers.} & \textbf{Payment} & \textbf{Search} & \textbf{\mini} & \textbf{\op} \\ \hline
01 & AliPay & \bluec & \bluec & \bluec & \bluec & \bluec & \redx & \bluec & \bluec & \bluec & \bluec & \redx & \redx \\
02 & Taobao & \bluec & \redx & \bluec & \bluec & \redx & \redx & \bluec & \bluec & \bluec & \bluec & \redx & \redx \\
03 & UC Browser & \bluec & \redx & \bluec & \bluec & \redx & \redx & \bluec & \bluec & \redx & \bluec & \redx & \redx \\
04 & Gaode & \bluec & \bluec & \redx & \redx & \bluec & \redx & \bluec & \bluec & \bluec & \bluec & \redx & \redx \\
05 & DingTalk & \bluec & \bluec & \bluec & \bluec & \bluec & \bluec & \bluec & \bluec & \bluec & \redx & \redx & \redx \\
06 & Youku & \bluec & \redx & \bluec & \redx & \bluec & \redx & \bluec & \bluec & \bluec & \bluec & \redx & \redx \\
07 & WeChat & \bluec & \bluec & \bluec & \bluec & \bluec & \redx & \bluec & \bluec & \redx & \redx & \bluec & \redx \\
08 & WeCom & \bluec & \bluec & \bluec & \bluec & \bluec & \bluec & \bluec & \bluec & \redx & \redx & \redx & \redx \\
09 & QQ & \bluec & \bluec & \bluec & \bluec & \bluec & \redx & \redx & \bluec & \redx & \redx & \redx & \redx \\
10 & QQ Music & \bluec & \redx & \redx & \redx & \bluec & \redx & \bluec & \bluec & \bluec & \bluec & \redx & \redx \\
11 & Tencent Video & \bluec & \redx & \redx & \redx & \redx & \redx & \bluec & \bluec & \bluec & \bluec & \redx & \redx \\
12 & TikTok & \bluec & \bluec & \bluec & \bluec & \bluec & \redx & \bluec & \bluec & \bluec & \redx & \redx & \redx \\
13 & Toutiao & \bluec & \bluec & \redx & \redx & \redx & \redx & \bluec & \bluec & \bluec & \bluec & \redx & \redx \\
14 & Feishu & \bluec & \bluec & \redx & \redx & \redx & \bluec & \bluec & \bluec & \bluec & \bluec & \redx & \redx \\
15 & Meituan & \bluec & \bluec & \bluec & \redx & \bluec & \bluec & \bluec & \bluec & \bluec & \bluec & \redx & \redx \\
16 & Dianping & \bluec & \redx & \redx & \redx & \redx & \redx & \bluec & \bluec & \bluec & \bluec & \redx & \redx \\
17 & Baidu & \bluec & \bluec & \bluec & \bluec & \bluec & \bluec & \bluec & \bluec & \redx & \redx & \redx & \redx \\
18 & Baidu Map & \bluec & \bluec & \redx & \redx & \bluec & \redx & \bluec & \bluec & \bluec & \bluec & \redx & \redx \\
19 & iQiYi & \bluec & \bluec & \bluec & \bluec & \bluec & \bluec & \bluec & \bluec & \bluec & \bluec & \redx & \redx \\
20 & PinDuoDuo & \bluec & \bluec & \bluec & \redx & \redx & \redx & \bluec & \bluec & \bluec & \bluec & \redx & \redx \\
21 & XiaoHongShu & \bluec & \bluec & \bluec & \bluec & \bluec & \redx & \bluec & \bluec & \bluec & \bluec & \redx & \redx \\
22 & KuaiShou & \bluec & \bluec & \bluec & \bluec & \bluec & \redx & \bluec & \bluec & \bluec & \redx & \redx & \redx \\
23 & NetEase Cloud Music & \bluec & \bluec & \bluec & \bluec & \bluec & \bluec & \redx & \redx & \redx & \redx & \redx & \redx \\
24 & JingDong & \bluec & \redx & \bluec & \bluec & \bluec & \bluec & \bluec & \bluec & \bluec & \bluec & \redx & \redx \\
25 & Suning & \bluec & \redx & \bluec & \bluec & \bluec & \redx & \bluec & \bluec & \bluec & \bluec & \redx & \redx \\
26 & Bilibili & \bluec & \bluec & \bluec & \bluec & \bluec & \bluec & \bluec & \bluec & \bluec & \bluec & \redx & \redx \\
27 & Grab & \bluec & \redx & \bluec & \redx & \redx & \redx & \bluec & \bluec & \bluec & \redx & \redx & \redx \\
28 & Paytm & \bluec & \redx & \redx & \redx & \redx & \redx & \bluec & \bluec & \bluec & \redx & \redx & \redx \\
29 & Go-Jek & \bluec & \bluec & \redx & \redx & \redx & \redx & \bluec & \bluec & \bluec & \redx & \redx & \redx \\
30 & UnionPay & \bluec & \bluec & \bluec & \redx & \redx & \redx & \bluec & \bluec & \bluec & \redx & \redx & \redx \\ 
31 & Air Asia & \bluec & \redx &  \bluec &  \bluec &  \bluec &  \bluec &  \bluec &  \bluec &  \bluec &  \bluec & \redx & \redx \\ \hline
& \textbf{Total} & \textbf{31} & \textbf{20} & \textbf{22} & \textbf{17} & \textbf{20} & \textbf{10} & \textbf{29} & \textbf{30} & \textbf{25} & \textbf{19} & \textbf{1} & \textbf{0} \\
\hline
\end{tabular}
\end{adjustbox}
\vspace{-0em}
\end{table*}

\noindent \textbf{Result.}
Over the past few years, there has been a significant focus on privacy protection.
We display the papers and detailed privacy data types in Table~\ref{tab:paper_summary}. 
We find that \textit{none} of the existing papers have studied privacy issues of the \mini and \op in super-apps. The privacy content that existing papers have focused on includes contacts, location, personal information (Pers.Info.), device information (Devi.Info.), and data generated by cameras, microphones, and sensors. Specifically, 22 papers concern personal information, such as phone numbers, ID numbers, gender, and age. 17 papers examine the protection of device information such as IMEI and MAC addresses, and 15 papers focus on system-level location information. 
Furthermore, increasing studies focus on broader aspects of privacy security. Nine and eight papers study the
potential privacy leaks from camera and microphone data, respectively. Four papers study the security risks of sensor data, as the adversary can use these data to infer user privacy attributes. These studies
underscore the multifaceted nature of privacy concerns on mobile devices and
reflect ongoing efforts within the academic community to address these
evolving challenges. Nevertheless, existing literature lacks a rigorous study on
the potential privacy leakage of \mini and \op.

\subsection{Practice in Industrial Companies}
\label{sec:industrial_practice}

\textbf{Methodology.}
To comprehensively analyze industrial companies' attitudes toward the types of private data, we study the most popular super-apps in the Google Play and Apple App Store. 
Our study includes two steps. First, we identified 31 different super-apps from recent research~\cite{yang2022cross,10.5555/3620237.3620608,zhang2022identity}. Second, we manually reviewed the privacy policies and terms of each super-app and summarized the types of privacy data in the privacy policies and terms. We regard the data types explicitly mentioned in the privacy policies and terms as the data the super-apps are concerned with. \revised{Consistent with the methodology in Section~\ref{sec:academia_community}, the four authors independently analyzed the privacy policies and terms, achieving a high IRR with a 99.93\% agreement rate.}

\noindent \textbf{Result.}
Table~\ref{tab:superapp_summary} displays the fine-grained categorization of private data that appear in the 31 super-apps' privacy policies and terms. We display twelve different types of privacy data,
including location, contacts, camera, gallery, microphone,
calendar, device information (Devi.),
personal information (Pers.), payment details, search history, \mini and \op.

We found that the industry does not pay enough attention to
\mini and \op. 
Among all super-apps, only one super-app (WeChat)
mentions that it collects \mini in the privacy policies and terms. \mini is
crucial for understanding users' behavior and preferences, as it includes the
mini-apps the user prefers. However, most industrial companies do not
consider \mini a primary privacy concern due to the lack of research on
its privacy leakage. From a company's perspective, \mini consists of structural elements, like the
types of mini-apps. Intuitively, these elements are predetermined and do not involve
user input data. 
Similarly, companies consider \op non-invasive because it only
captures limited structured behavioral data without explicitly requiring personal input.

Conversely, super-apps often have clear notifications on other
common privacy data. Personal information and location data are commonly
included in privacy policies and terms, highlighting their importance for app
functionality. Device information is also frequently collected for security and
optimization. Payment data helps the super-app to understand user
preferences. Additionally, 22, 17, and 20 super-apps claim access to cameras, galleries,
and microphones. Contacts and calendar data are collected by 20 and 10 social-based super-apps, respectively. Search history is widely used in 19 super-apps, raising attention among users for its sensitive content.

\section{Threat Model}

\revised{
Given people's unawareness of the security risk of mini-app interaction history, the next part of this paper will reveal the dangers. In this section, we will first introduce the threat model in which the interaction data can be used to threaten users' privacy.
}

\noindent \textbf{Scenario.}
We consider a leading global super-app vendor that consists of multiple departments, each with distinct roles and responsibilities. 
In this scenario, there are two types of departments: the high-privilege business department ($D_{high\_priv}$), such as payment and risk control, and the low-privilege business department ($D_{low\_priv}$), such as lifestyle or third party-collaborators. Both  $D_{high\_priv}$ and $D_{low\_priv}$ strictly comply with privacy protection measures to ensure service security and regulatory compliance. $D_{high\_priv}$ has access to sensitive data, such as personal financial information, while $D_{low\_priv}$ can only access data explicitly classified as non-sensitive by the super-app vendor and should know minimal user privacy. 

\revised{In this scenario, the insider threats~\cite{6957307} within the $D_{low\_priv}$ may act as the adversary. These insiders cannot access private data but can access mini-app interaction history because it is considered non-sensitive. For instance, employees in advertising or loan departments cannot directly access sensitive user attributes due to data-isolation policies. However, they can access \mini and \op to infer those attributes indirectly. Their motivation may include selling the inferred data for profit and gaining a competitive advantage by enhancing targeted advertising or differential pricing strategy. Even when their mini-app does not hold explicit data (e.g., a user’s assets status), relevant privacy attributes of unlabeled users can still be inferred by analyzing broader interaction patterns. As an example, identifying whether a user owns property or a vehicle helps loan department insiders tailor services more effectively. Meanwhile, since users rarely access the loan department’s mini-app, the department cannot gather enough privacy attributes directly. Nonetheless, these attributes can still be inferred by analyzing the mini-app interaction history.}

\noindent \textbf{Adversary's Goal.}
\revised{The adversary's goal is to \textit{accurately predict the privacy attributes of as many users as possible from the non-sensitive data}.}
The goal consists of twofold. 
First, the adversary aims to identify a user subset whose privacy attributes are strongly correlated with \mini and \op. These users are believed to have sufficient mini-app interaction history data to
infer their privacy attributes. 
For example, one user might frequently use the fueling mini-apps, which could indicate that he/she owns a vehicle.
For other users, if the \mini and \op cannot reveal their privacy attributes, the adversary may classify them as unknown rather than forcibly assign labels. This can avoid blind and inaccurate predictions.
Second, for users in the subset, the adversary aims to infer their privacy attributes as accurately as possible. This is because privacy attributes are valuable but scarce. With such attributes, adversaries can design unfair market strategies (e.g., targeted advertising, price discrimination, or denial of services) for certain user groups. For example, they might use financial status to implement differential pricing that targets wealthier users with higher service fees. Meanwhile, all super-app users are potential victims because many users do not provide such information due to privacy concerns. For instance, according to \alipay's data, only 6.2\%, 5.8\%, and 6.9\% users provide their marital status, property ownership, and vehicle ownership, respectively.

Note that this goal is
realistic and dangerous because super-app vendors often have a large
user base (over one billion). Even if the adversary can only infer the privacy of
a small portion of users, the number of threatened users is significant. Although the super-app vendor strives to control access by $D_{low\_priv}$, as super-app functionality expands, data initially considered privacy-irrelevant may become more privacy-revealing, leading to new risks.

\noindent \textbf{Adversary's Ability.}
\revised{Unlike prior work~\cite{tu2018your} that requests a vast number of OS-level data, we assume the adversary can acquire only \mini and
\op from the vendor's server because existing super-apps (\eg Alipay, WeChat, and Tiktok) are web-based and hosted on the servers~\cite{alipaysdk,wechatsdk,tiktoksdk}, the vendor naturally needs \mini and \op to provide the desired app functionality.} In practice, the privilege level of different data is determined by regulations, industrial practices, or research papers. Since \mini and \op are not identified as sensitive in these above sources, as discussed in Section~\ref{sec:privacy_of_history}, they are not identified as sensitive and can be accessed by $D_{low\_priv}$.

Due to regulation and performance constraints, we assume the attacker cannot modify the released versions of the super-apps on users' devices, such as instrumentation or probing, to collect user information on the device side.

We also assume that the adversary can collect the privacy attributes of a small number of users from a publicly available data breach. These breaches are widely available on the Internet~\cite{liu2015cloudy,mayer2021now,thomas2017data}, which contains phone numbers, personal identification numbers, and other sensitive privacy attributes. The adversary can match their user data to the samples in breach databases (e.g., by matching phone numbers~\cite{cheng2017enterprise, thomas2017data}) to identify the privacy attributes of a few hundred users. The adversary can then collect \mini and \op of these users to train the model to infer the privacy of other users.

\section{THEFT: Mini-app Interaction History-Oriented Inference Attack}

\revised{
To reveal the new security risk of mini-app, we design a new attack, \tool, \textit{mini-app interac\underline{T}ion \underline{H}istory-ori\underline{E}nted in\underline{F}erence a\underline{T}tack}. This attack is proof of the potential security issue of the interaction data.}
\tool uses both \mini and \op to infer the privacy attributes of users. We will first illustrate the overall pipeline, followed by a detailed presentation of each component.

\subsection{Overview}

Generally, the \tool employs a DNN model to infer user privacy. The inputs of the DNN model are \mini and \op, and the model outputs are the predicted attribute label and a confidence score, which is used to filter out the high-confidence victim subset. The confidence score needs to approximate the real accuracy of the prediction. For example, if a model predicts the user to be a female with a confidence level above 0.9, then the accuracy of this prediction should be more than 90\%. Using a predefined threshold, the adversary can select high-confident samples to identify a vulnerable victim subset and provide label predictions while labeling other users with lower confidence as unknown.

\begin{figure*}[!t]
    \centering
    \includegraphics[width=0.95\linewidth]{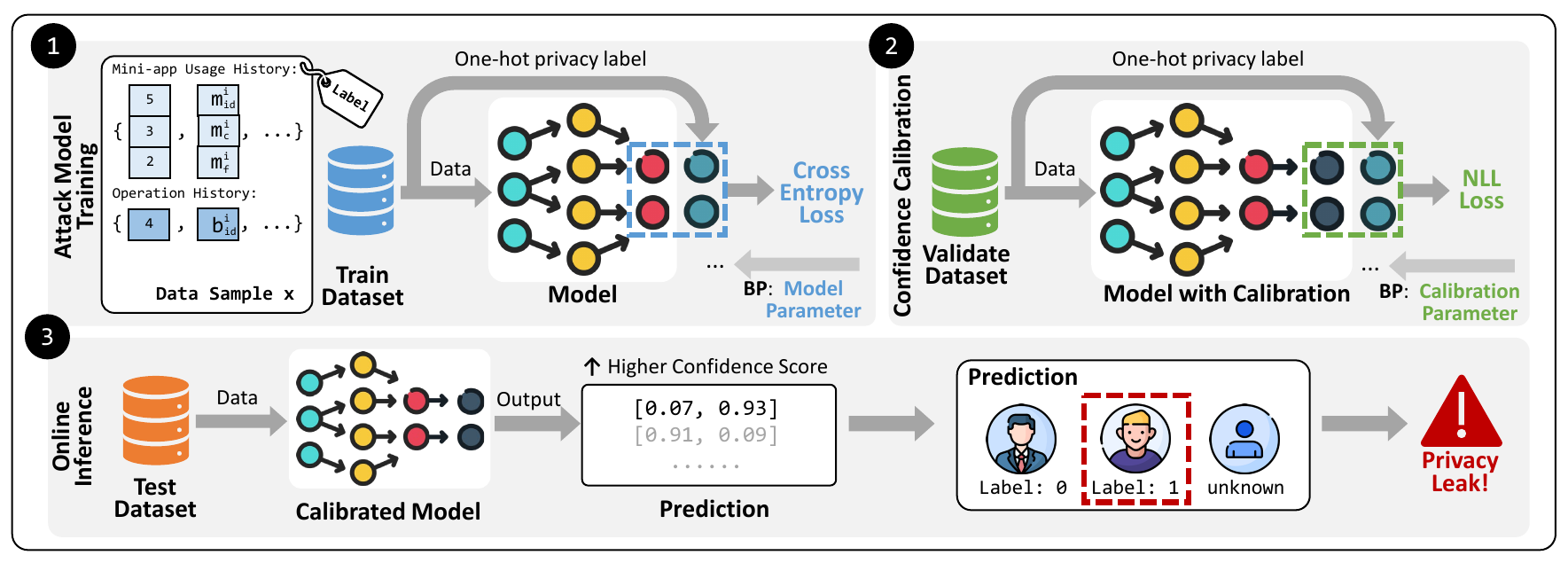}
    \vspace{-1em}
	\caption{The pipeline of \tool consists of three steps: attack model training, confidence calibration, and online inference.}
	\label{fig:attack}
 \vspace{-1em}
\end{figure*}

The attack process consists of three stages, as shown in
Figure~\ref{fig:attack}. The first step is the attack model training, where the
adversary uses the \mini and \op data of a relatively small set of leaked users with one-hot privacy attribute labels to train a DNN
model.
The second step is model confidence calibration, where the adversary uses a
calibration technique to ensure that the confidence score can faithfully reflect
the accuracy of the prediction.
The last step is online inference, where the adversary uses the trained
model to predict the privacy attributes of unlabeled users from \mini and \op.

\subsection{Attack Model Training}

\begin{table}[!tb]
\centering
\caption{28 categories of mini-apps.}
\vspace{-1em}
\label{tab:category}
\begin{adjustbox}{max width = 1.\columnwidth}
\begin{tabular}{@{}cl|cl@{}}
\hline
\textbf{Code} & \textbf{Category} & \textbf{Code} & \textbf{Category} \\ \hline
1 & \texttt{Education} & 15 & \texttt{Finance} \\
2 & \texttt{Entertainment} & 16 & \texttt{Food and Drink} \\
3 & \texttt{House and Home} & 17 & \texttt{Health and Fitness} \\
4 & \texttt{Lifestyle} & 18 & \texttt{Art and Design} \\
5 & \texttt{Maps, Navigation, and Taxi} & 19 & \texttt{Books} \\
6 & \texttt{Music and Audio} & 20 & \texttt{Comics} \\
7 & \texttt{Parenting} & 21 & \texttt{Communication} \\
8 & \texttt{Shopping} & 22 & \texttt{Medical} \\
9 & \texttt{Auto and Vehicles} & 23 & \texttt{News} \\
10 & \texttt{Beauty} & 24 & \texttt{Photo} \\
11 & \texttt{Business} & 25 & \texttt{Productivity} \\
12 & \texttt{Dating} & 26 & \texttt{Sports} \\
13 & \texttt{Social} & 27 & \texttt{Weather} \\
14 & \texttt{Travel and Local} & 28 & \texttt{Event} \\ \hline
\end{tabular}
\end{adjustbox}
\vspace{-1em}
\end{table}

In the attack model training stage, \tool trains a model that can better
learn the features of mini-app interaction history data and thus can accurately
predict the privacy attributes of users. The training data are the mini-app
interaction history (\mini and \op), the training labels are
collected privacy attributes from data breaches. We select the
transformer-based~\cite{vaswani2017attention} model as the default architecture
of \tool. The transformer model comprises 12 encoders, each equipped with
12 bidirectional self-attention heads, resulting in a total of 110 million parameters. In Section~\ref{sec:architecture_comparison}, we also compare our transformer
architecture with another two architectures to demonstrate the superiority of
our choice. For each type of privacy attribute, we modified the output dimension of
the last fully connected layer to the number of labels.

\noindent \textbf{Model Input.}
The model input consists of two parts, \mini (denoted as $x_m$) and \op
(denoted as $x_o$). $x_m$ contains the list of the latest $N$ mini-apps accessed
by a user. For the $i$-th mini-app in the list, we record 1) the unique id
$m^{i}_{id}$ of the mini-app (maintained by the \alipay backend), 2) the
mini-app category code $m^i_{c}$, and 3) the number of access times $m^i_{f}$
over the last 30 days. For the category code of mini-apps, we classify the
mini-apps into 28 types following prior literature~\cite{li2022collect}. Table~\ref{tab:category} displays all the categories. Therefore, $x_m$ is a tensor
of the shape of $(M, 3)$. 
Meanwhile, $x_o$ contains button IDs that the user clicked over the past $M$ timestamps, where each timestamp represents a 500-millisecond interval. \revised{We convert the original click logs with timestamps from \alipay into this format, to unify the input format for our deep learning model.}
Within each interval, if the
user clicks a button in the mini-app, we record its ID. Otherwise,
we record $0$ to indicate that the user does not click any button. These button IDs are globally unique in the super-app, with each ID corresponding to a specific function and remaining consistent across all mini-apps. The input of the model is the fusion of
$x_m$ and $x_o$. To fuse the two types of data, we empirically set $N=M$ and
concatenate $x_m$ and $x_o$ into a single input tensor for the model.

\noindent \textbf{Model Output.}
The output is the predicted label for a specific type of privacy attribute. In this
paper, we focus on seven privacy attributes: gender, location, age, property
ownership, vehicle ownership, marital status, and parental status. All these
types of privacy are maintained by \alipay and are considered important privacy
information in the internal company. For each privacy attribute, we regard it
as a classification task. For five privacy attributes, the inference task is a
binary classification task. The five privacy attributes are gender, whose labels are
male and female, and the ownership of property and vehicles, marital and
parental status, whose labels are yes and no. For location, we give three
labels: Tier-1 cities, Tier-2 cities, and Tier-3 cities, making this task a ternary
classification task. \revised{Specifically, Tier-1 cities include municipalities, well-developed provincial capitals, and economic centers; Tier-2 cities encompass other provincial capitals and major cities; and Tier-3 cities comprise ordinary cities and regions primarily composed of towns and villages.} Regarding age, we follow prior research and assign four labels:
Under 18, 18$\sim$39, 40$\sim$65, and Above 65~\cite{shauly2020public}.

\subsection{Model Confidence Calibration} 

\textbf{Necessity.}
Recall that the goal of \tool is to identify a subset of users whose
privacy attributes can be inferred accurately. However, the original
output confidence of the model cannot faithfully reflect the
inference accuracy. This is because the original model is optimized by the
cross-entropy (CE) loss. The CE loss will lead to overconfidence because it greedily maximizes
the confidence of the predicted label. However, for samples in which the model
is not confident, we do not want to give a high confidence score to
the predicted label because it will lead to a high false positive rate. In other
words, when the attack model generates a confidence score of 0.9 for the
prediction of user privacy, it should also ensure that the prediction accuracy
is above 0.9. Therefore, by selecting predictions with high confidence scores,
we can identify the subset of users from which we can accurately infer their
privacy attributes.

\noindent \textbf{Temperature Calibration.}
We adopt model calibration techniques~\cite{muller2019does,wang2021rethinking}
to achieve our goal. The general idea of model calibration is to append
an extra trainable calibration module after DNN's output layer. 
The calibration module takes DNN's output (overconfident
prediction) as input and generates a new confidence score. The calibration
module is fine-tuned on a calibration dataset to minimize the gap between the
generated confidence score and the real accuracy. Our design chooses the
temperature calibration~\cite{muller2019does,wang2023tabi} because it performs best in our evaluation. In Section~\ref{sec:ablation_study}, we
compare our temperature calibration with two other calibration techniques to
demonstrate the superiority of our choice.

Specifically, given a trained attack model, we replace its softmax function with
a calibrated softmax function, which is shown in Equation~\ref{eq:cali_tem}. The
calibrated softmax adds a trainable parameter, temperature $t$, to the softmax
function. Then we fine-tune $t$ by minimizing the negative log-likelihood loss
(NLL loss) in the validation dataset. Note that we fixed the parameters of the
attack model during the calibration phase and only updated $t$. The loss function choice and the validation dataset's use are consistent
with prior work~\cite{pmlr-v80-kuleshov18a,wang2023tabi}. 
\begin{equation}
    C(x) =\frac{\max_{i \in K}\exp(o_i / t)}{\sum_{i \in K} \exp(o_i / t)}
    \label{eq:cali_tem}
\end{equation}

\subsection{Online Inference} 

For inference, we use a pre-defined threshold $t_d$ as the confidence bar to
determine whether our model can confidently infer the privacy attribute of a
user. For the input of a user $x$, the output confidence score of the calibrated
module is $C(x)$. \tool returns ``unknown'' if $C(x)$ is smaller than $t_d$.
Otherwise, \tool will generate the corresponding attribute label for the given user.

\section{Evaluation}

In this section, we comprehensively evaluate the inference performance of \tool. 
We first describe the experimental setup in Section~\ref{sec:exp_setup}. 
In Section~\ref{sec:attack_effectiveness}, we
display the effectiveness of \tool, particularly the high inference accuracy
on the high-confidence samples. After that, in Section~\ref{sec:exp_insights},
we provide insight and in-depth analysis on the success
of \tool for each of the privacy attributes. In Section~\ref{sec:architecture_comparison}, we prove the effectiveness of our model architecture selection.
In Section~\ref{sec:ablation_study}, we conduct ablation studies to
demonstrate the generalizability of our approach and the recommended parameter
settings.

\noindent\revised{\textbf{Ethical Disclaimer. }Since our evaluation involves collecting privacy attributes and mini-app interaction history, we obtained approval from the IRB of \alipay. For details, please refer to the Ethics Considerations Section.}

\subsection{Experiment Setup}
\label{sec:exp_setup}

\textbf{Volunteer Selection.}
We first select 5\% of daily active users of \alipay as volunteer candidates and
send invitations to them. To guarantee the quality of the data, we asked the
internal engineers of \alipay to confirm that the candidates have been active in
\alipay for more than 7 days at the time of invitation. In each invitation, we
provided a clear consent notice on what information is collected and stated
that the data were collected only for research purposes. The candidates who
accepted the invitation become our participants. In total, we selected
38,351,206 candidates, of which 288,895 users agreed to share their data.

For our participants, after their consent, we pushed the \alipay-dev version to their devices, replacing the original version. The
\alipay-dev version is the same as the original version. The only difference is that it displays an additional user agreement when the user first opens the app. This agreement notifies the users that he/she is involved in our experiment and displays detailed information about our experiments. Note that we do not add extra instrumentation or probes in \alipay-dev. We store \mini and \op in an encrypted database to protect the privacy of volunteers.

\noindent \textbf{Data Collection.}
We collect data during the whole process of using \alipay. Specifically, the
collection phase starts when the user opens the app and ends when the user exits
\alipay or switches to another app. We collect both \mini and \op during this phase. We collected data for 50 days and obtained 10,987,662 data samples from 288,895 users.

We empirically set the length $N$ to 200 for each mini-app interaction history data sample. Specifically, for \mini, we record the Top 200 recently used mini-apps. We confirmed with the internal engineers of \alipay that 200 mini-apps
are sufficient to represent the user's recent behavioral patterns, as the
average number of mini-apps used by \alipay users is 117.8.
\revised{For \op, we convert the original logs in \alipay with timestamps into 0.5-second intervals over a 100-second window, resulting in a 200-dimensional vector per data sample.}
We empirically determined these values based on \alipay's report that the average time of user interactions is 89.2 seconds. A period of 100 seconds is sufficient to cover most
user interactions. The internal engineers of \alipay also confirmed that the
500-millisecond interval is the minimum for user interactions, and most
interactions are shorter than 100 seconds. Recent literature also supports the setting of these values~\cite{10.1145/2750858.2805837,10.1145/3604271}. Therefore, the
dimension of each mini-app interaction history data sample is $200 \times 4$.

\noindent \textbf{Building the Ground Truth.} 
We strictly adhered to IRB requirements (illustrated in the Ethics Considerations
Section) to build ground truth privacy attributes as training labels.
For the location attribute, we obtained user consent to access the data of the system
location service. For other attributes, we first ask users to provide their
information. Then we sent authorization requests to acquire user consent to
access users' data in \alipay's server, thereby obtaining
gender, age, as well as property and vehicle information registered under their
names to confirm user-provided labels. 
For users who are unwilling to share all
their privacy attributes or the shared attributes do not match the data in \alipay's server,
we removed their data in our evaluation. In summary, we construct the dataset with 1,099,130 data samples from 219,826 users.

\noindent\textbf{\revised{Unbiasedness of Collected Dataset.}}
\revised{We further demonstrate that the collected dataset closely aligns with real-world population distributions.}
We analyzed label distribution for various privacy attributes: 22.8\% users are from tier-1 cities, 43.4\% from tier-2, and 33.8\% from tier-3. The gender distribution is almost equal: 50.3\% male and 49.7\% female. For age, 17.5\% are under 18, 38.0\% are between 18-40, 30.9\% are 40-65, and 13.6\% are above 65. Moreover, 48.6\% users own property, 21.0\% users own vehicles, 69.7\% are married, and 53.6\% have children. To verify the unbiasedness of our data, we conducted a chi-square test comparing these distributions with data from China’s National Bureau of Statistics~\cite{NBSC}. \revised{The test confirmed no significant statistical difference, indicating that our dataset is indeed representative of the wider population.}

\noindent \textbf{Experiment Protocol.} 
We split the collected data into three sets: a training set, a validation set, and a test set. To demonstrate that \tool can use a very small dataset as the training set to infer the privacy attributes of large-scale users, we set the size of training and validation sets to a small number (less than 0.1\% of the test set). Specifically, the training set includes 200 users, and the validation set includes another 200 users. We collect five data samples from each user. Thus, the total number of training and validation sets is 1,000. The test set contains 1,097,130 data samples from the other 219,426 users.
Furthermore, we repeated the evaluation ten times with random data partitioning to provide a reliable assessment.
For all comparisons in our evaluation, we run the hypothesis test~\cite{chen2023obsan} to ensure the significance of our observation.

\noindent \textbf{Implementation Details}
We conducted our experiment on a server with an Intel Xeon E5-2678 v3 CPU (48
cores), 128GB RAM, and 2 NVIDIA RTX 3090 GPUs. The server OS is Ubuntu 20.04
LTS. We implemented our code base using Python 3.8 and PyTorch 2.0. All models are trained for 100 epochs. 
We monitor the loss in the validation set and ensure that the models are converged.
We set the discriminator threshold ($t_d$) to 0.9, meaning we identify data samples with a confidence score greater than 90.0\% as a high-confidence subset, according to the literature~\cite{kuppers2020multivariate,wang2021confident}. 
We utilize Adam as the optimizer, and the learning rate is set to 1e-3. 

\subsection{Attack Effectiveness}
\label{sec:attack_effectiveness}

In this section, we will first introduce the general effectiveness of \tool. Then, we will show that
\tool can effectively select a subset of users and accurately infer their
privacy attributes. Finally, we will illustrate that output confidence
of our model is aligned with the true accuracy of the prediction,
which is an essential indicator of the attack's effectiveness.

\begin{figure*}[!t]
    \centering
    \includegraphics[width=1.\linewidth]{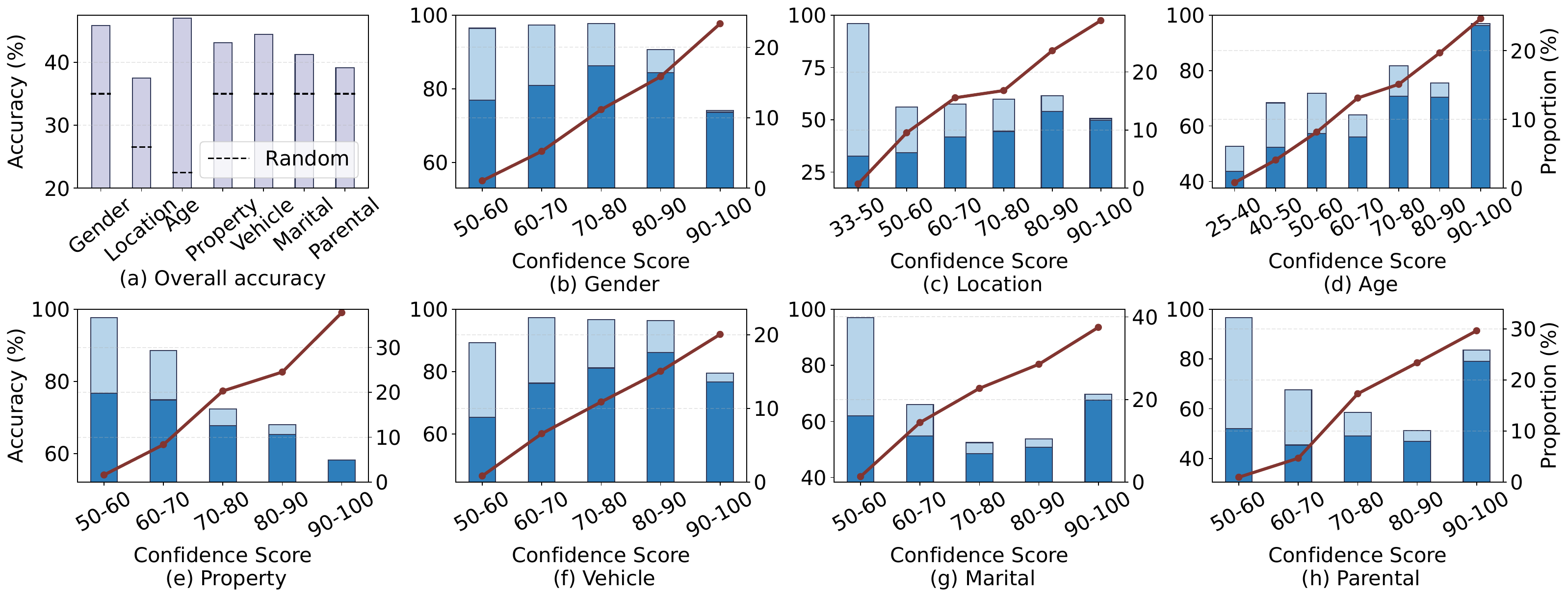}
    \vspace{-1.5em}
	\caption{Attack effectiveness of \tool. (a) Overall inference accuracy for each type of privacy. (b)-(h) For each privacy type and confidence interval, the proportion of users predicted to be within the interval ($P_{int}$), the proportion of correctly predicted users ($P_{conf}$), and inference accuracy ($acc_{int} = P_{conf} / P_{int} \times 100\%$).}
	\label{fig:results_all}
 \vspace{-1em}
\end{figure*}

\noindent \textbf{General Effectiveness.}
\revised{In Figure~\ref{fig:results_all}~(a) we show the overall inference accuracy across all inference samples for
each privacy attribute label.} 
We ran each experiment 10 times and reported the average
 accuracy. 
We manually checked the standard error and confirmed that for all
labels, the standard error is below $0.7\%$. Therefore, we omit the standard
error in the figure. In Figure~\ref{fig:results_all}~(a), we also mark the accuracy of the random
guess baseline with a dashed line. The average inference accuracy of \tool
is $65.2\%$, which is $48.1\%$ higher than the baseline. In particular, for the age of users, \tool is $2.9\times$ higher than the baseline. 
We also performed a chi-square test to confirm the superiority of \tool. Our
null hypothesis is that the performance of the model is equal to random guessing, and the $p$ value is $0.02$. Therefore, we reject the null hypothesis ($p < 0.05$) and conclude that \tool is significantly better
than the baseline.

\noindent \textbf{Performance on High-Confident Subset.}
In Figure~\ref{fig:results_all} (b)-(h), we split the model confidence score
into intervals of 10\%. Each subfigure represents the results of one type of
privacy. For each confidence interval, we show the proportion of samples within
this interval ($P_{int}$, represented in light blue bars) and the
correctly predicted samples within that interval ($P_{conf}$, represented in
dark blue bars). Note that the more dark blue bars cover the light blue bars,
the more samples of $P_{int}$ are correctly predicted (a higher
attack performance). We also show the prediction accuracy within that interval
($acc_{int} = P_{conf} / P_{int} \times 100\%$, represented by the red line).
For each subfigure, the y-axis on the left represents the prediction accuracy
($acc_{int}$), and the right y-axis represents the proportion of samples within
the interval ($P_{int}$ and $P_{conf}$).

For samples predicted by our model with high confidence scores, the
attack performance is \textit{near-perfect}. Setting the confidence
threshold as 90\%, we can on average identify 16.1\% samples, and the accuracy of the inference is 95.5\%. For the bar at the index 90-100 in
Figure~\ref{fig:results_all}~(b)-(h), the dark blue bars cover almost entirely the light blue bars. which means that the proportion of correctly predicted samples
($P_{conf}$) is very close to all samples in this interval ($P_{int}$). This
implies that our attacks are very likely to be successful in 15.4\% of all data
samples.

We can also observe that high attack performance is consistent across all
types of privacy. For six out of seven types of privacy, we can identify more than 10\% users with an accuracy of more than 90\%. For gender, we can identify
11.0\% data samples with an accuracy of 97.7\% by setting the confidence
threshold to 90\%. For location and age, we can identify 12.0\% and 23.9\% data
samples with an accuracy of 97.3\% and 98.6\%, respectively. For property
ownership, we can identify 4.9\% data samples with an accuracy of 99.0\%. For
vehicle ownership, we can identify 14.7\% data samples with an accuracy of
91.9\%. For marital status, we can identify 19.9\% data samples with an accuracy
of 93.5\%. For parental status, we can identify 25.8\% data samples with an
accuracy of 91.3\%.

\revised{Besides, our analysis shows that the distribution of users with high prediction accuracy is demographically representative of the overall collected dataset, indicating no bias toward specific groups. We perform a chi-square test to check consistency. The results show that the distributions of high-confidence data (p-value = 0.87) and other data (p-value = 0.96) align with the distribution of the entire test dataset, further confirming the absence of bias.}

\noindent \textbf{Correlation between Confidence and Accuracy.}
From the red lines in Figure~\ref{fig:results_all} (b)-(h), we can also observe
that \textit{the confidence score produced by our model is positively correlated
with accuracy}. We calculate the Pearson's correlation coefficient between confidence and accuracy, which is $0.992$ on average. It implies that for a given data sample, the higher the
confidence score, the more likely the prediction is to be correct. This is an
important indicator of the effectiveness of the attack. It means that we can
effectively select the subset of users by model confidence, and for the selected
users, we can accurately infer their privacy attributes.

\subsection{Insights From Data}
\label{sec:exp_insights}

To further understand the reasons for the privacy leakage, we
conduct a thorough investigation into the users that the model has high
confidence scores to identify the sources of leakage for each
attribute. 
We found that the users' privacy attributes relate to the preferred mini-app types and how they operate smartphones.
Specifically, we mainly study three metrics: two metrics are related to \mini: the number of mini-apps (\#Mini-app) and the access frequencies (\#Access) of specific mini-app types, and one metric relates to \op: the number of button clicks (\#Click) in each \op sample.

\begin{figure}[!t]
    \centering
    \includegraphics[width=1.\columnwidth]{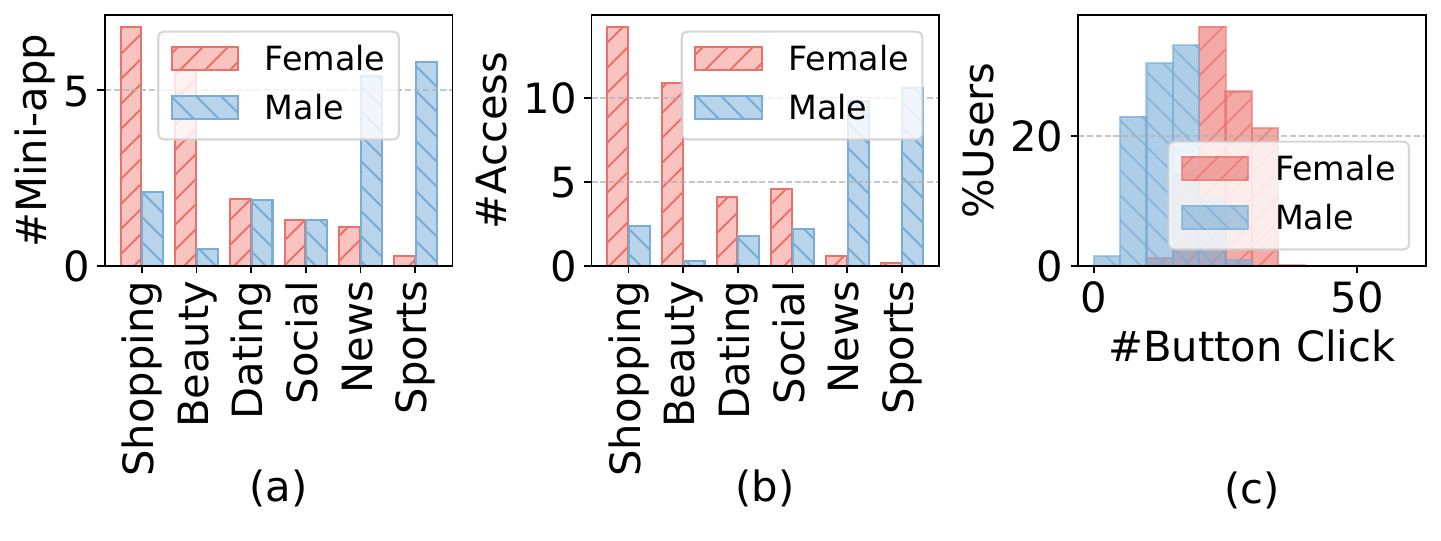}
    \vspace{-2em}
	\caption{Detailed analysis on the privacy of \textit{gender}.}
	\label{fig:gender}
\end{figure}

\noindent \textbf{Gender.}
We display \#Mini-app, \#Access of representative mini-apps in Figure~\ref{fig:gender}~(a) and (b). In
Figure~\ref{fig:gender} (c), we plot the user distribution w.r.t. \#Click. For all figures, the red bars represent female
users, and the blue bars represent male users.
For \mini, we
found that female users prefer \texttt{Shopping} and
\texttt{Beauty} mini-apps. Female users averagely use $6.8$ and
$5.9$ kinds of mini-apps for \texttt{Shopping} and \texttt{Beauty}, with a frequency
of $14.2$ and $10.9$ times. Conversely, male users only access
$2.1$ \texttt{Shopping} mini-apps for $2.4$ times, and $0.5$ \texttt{Beauty} mini-apps
for $0.3$ times. 
We found that male users prefer \texttt{News} ($5.4$ mini-apps for $9.8$ times) and
\texttt{Sports} ($5.8$ mini-apps for $10.6$ times). Conversely, female users
only access $0.7$ mini-apps of these two categories for $0.4$ times.
For \op, female users usually operate more frequently than male
users, confirmed by a recent study~\cite{tian2021and}.
Figure~\ref{fig:gender}~(c) shows that male's \#Click ranges from 10 to
20, while female's \#Click ranges from 20 to 35. This observation aligns
with recent research showing that females tend to use apps for a longer time than
males~\cite{tian2021and}.

\noindent \textbf{Location.} 
\mini and \op can reflect the economic and population status of
the city where users live, thus leaking the location attribute. To display
this result, we first show the statistical observation of \mini, then map the
mini-app access frequencies to the users' geographical locations in
China, and display the statistics of \#Click.

\begin{figure}[!t]
    \centering
    \includegraphics[width=0.95\columnwidth]{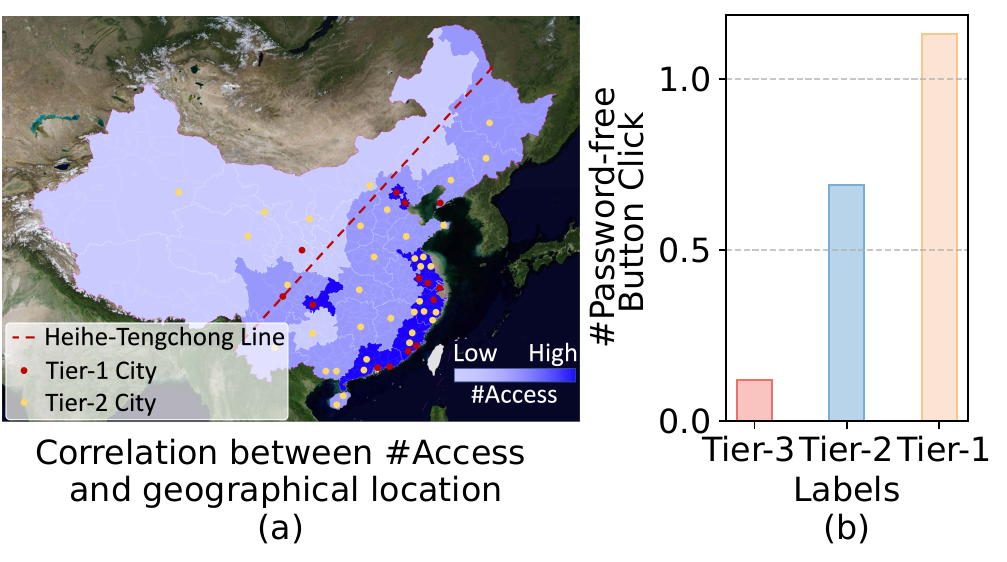}
	\caption{\revised{Detailed analysis on the privacy of \textit{location}.}}
	\label{fig:loc_map}
\end{figure}

For \mini, both \#Mini-app and \#Access correlate with users' location. Users in Tier-1, Tier-2, and Tier-3 cities averagely utilize 27.9, 23.6, and 17.3 types of mini-apps. For \#Access, Users in Tier-1, Tier-2, and Tier-3 cities use mini-apps for 36.3, 20.9, and 13.1 times, respectively.
To further verify our observation, we map \#Access to the
users' geographical locations in China (by province) in
Figure~\ref{fig:loc_map}~(a). The darker color represents a higher \#Access. Red and yellow dots represent the Tier-1 and Tier-2 cities. We plot the Heihe-Tengchong Line~\cite{guo2014scientific} with a dashed red line. Heihe-Tengchong Line is a
famous geographical line in China representing population density. The line's southeast side has 94\% of China's population and represents a higher economic level. The northwest side only has 6\% of the population and represents a lower economic level.
From Figure~\ref{fig:loc_map}~(a), we can see that the southeast (lower right in the figure) area of the line is darker than the northwest (upper left) area, which means \#Access is positively correlated with the economic level. We think it is because people in more developed cities rely more on digital services daily. For example, people in Tier-1 cities are more likely to use mini-apps to order food, buy tickets, and pay utility fees.

For \op, we observe a remarkable difference in \#Click between users from different locations. 
Specifically, we focus on the button of \texttt{password-free payment}~\cite{liu2015role}.
This button lets users execute small-amount transactions (below 200 RMB, approximately 30 USD) without inputting passwords. This button is designed to improve user experience and avoid frequently entering passwords. We found that users in the more developed areas have higher \#Click on \texttt{password-free payment}. For users from Tier-3, Tier-2, and Tier-1 cities, \#Click on \texttt{password-free payment} is $0.1$, $0.7$, and $1.1$, respectively. We also observe that this statistical result is specific to the \texttt{password-free payment} button. We do not observe this phenomenon on other buttons with similar functions, such as the \texttt{payment} button.

\noindent \textbf{Age.} 
The good performance of \tool for age is because
people of different ages display different patterns on the types of mini-apps and
how they interact with mini-apps. 

\begin{figure}[!t]
    \centering
    \includegraphics[width=1.\columnwidth]{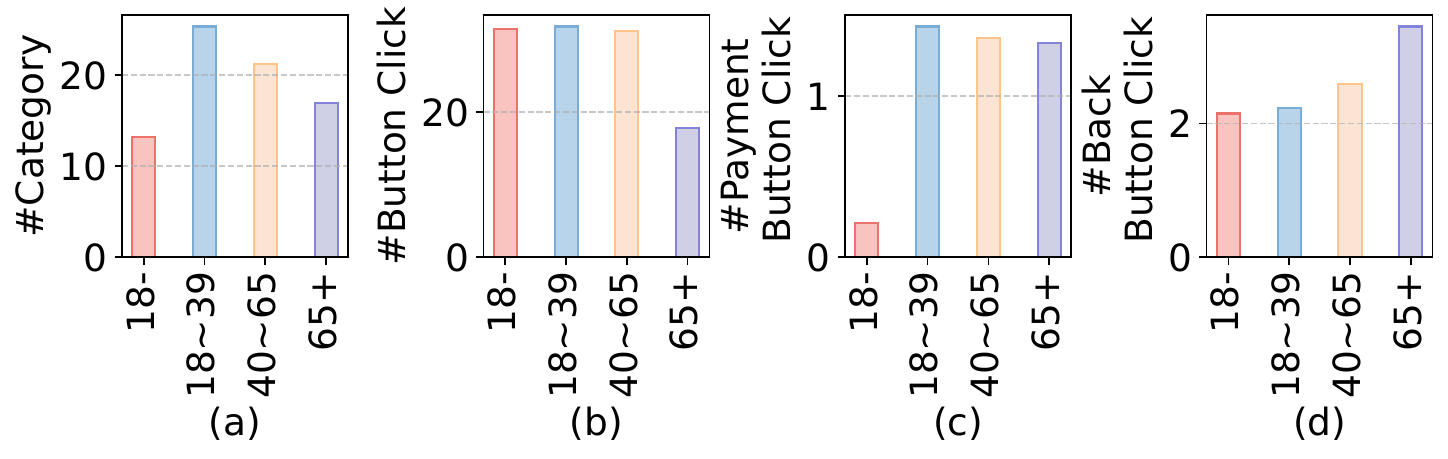}
    \vspace{-2em}
	\caption{Detailed analysis on the privacy of \textit{age}.}
	\label{fig:age}
\end{figure}

For \mini, we plot the number of accessed mini-app categories (\#Category) in
Figure~\ref{fig:age}~(a). Minors (\texttt{Under 18}) and the elderly
(\texttt{Above 65}) access a relatively smaller \#Category (13.2 and 16.9) than other age
groups (above 21.0). 
Notably, minors rarely access mini-apps of \texttt{Finance} because regulations restrict financial services to minors. Similarly, elderly users rarely access mini-apps of 
\texttt{Dating} and \texttt{Comics}, as
older people are not the target users of these mini-apps.

For \op, we display \#Click on any buttons (\#Button Click), \#Click on the
\texttt{Payment} button (\#Payment Button Click), and \#Click on
the \texttt{Back} button (\#Back Button Click) of each age group in
Figure~\ref{fig:age}~(b)-(d). 
For \#Button Click, elderly users have the lowest value (average $17.8$) compared to other groups ($31.5$).
This is because the elderly are less familiar with using super-apps and tend to react more slowly to the response.
For \#Payment Button Click, minors rarely click this button (average $0.2$ clicks) than other groups ($1.4$ clicks) because minors are not allowed to make payments without the consent of their guardians. 
For \#Back Button Click, the elderly have the highest value ($3.5$) compared to other groups ($2.3$) because the elderly usually cannot accurately select the desired button and thus need to click the \texttt{Back} button to return to the previous page and try again.

\begin{table}[t]
\centering
\caption{Comparison for users with and without property.}
\vspace{-0.5em}
\label{tab:property}
\begin{adjustbox}{max width = 0.9\columnwidth}
\begin{tabular}{@{}cccc@{}}
\hline
\textbf{Mini-app} & \textbf{Prodived} & \multicolumn{2}{c}{\textbf{\#Access}} \\ \cline{3-4} 
\textbf{Name} & \textbf{Services} & \textbf{W/o Prop.} & \textbf{W/ Prop.} \\ \hline
Utilities & Household bills payment & 0.1 & 2.6 \\
State Grid & Electricity bill payment & 0.0 & 1.3 \\
CSGrid & Electricity bill payment & 0.0 & 1.1 \\
Ziroom & Housing rental & 1.2 & 0.0 \\
Anjuke & Housing rental & 1.1 & 0.0 \\
Renting & Housing rental & 1.6 & 0.0 \\ \hline
\end{tabular}
\end{adjustbox}
\end{table}

\noindent \textbf{Property.} 
Table~\ref{tab:property} shows representative mini-apps, the services provided by each mini-app, and the access
frequencies of users with and without property. Users without any
property rarely access mini-apps that can pay household bills (\eg \texttt{Utilities}, \texttt{State Grid},
and \texttt{CSGrid}; frequency below $0.1$). Contrarily, users with property access these mini-apps more frequently (an average of $1.7$ times). This is because
these users must pay the household bills themselves.
On the other side, users without property prefer mini-apps about
house renting (\eg \texttt{Ziroom}, \texttt{Anjuke}, and \texttt{Renting}). The access frequency ($1.3$) is higher than users with property ($0.0$). This is because users without property are likelier to rent a
house.

\begin{table}[t]
\centering
\caption{Comparison for users with and without vehicles.}
\label{tab:vehicle}
\begin{adjustbox}{max width = 0.9\columnwidth}
\begin{tabular}{@{}cccc@{}}
\hline
\textbf{Mini-app} & \textbf{Prodived} & \multicolumn{2}{c}{\textbf{\#Access}} \\ \cline{3-4} 
\textbf{Name} & \textbf{Services} & \textbf{W/o Veh.} & \textbf{W/ Veh.} \\ \hline
12123 & Traffic police platform & 0.2 & 1.5 \\
Sinopec & Fueling & 0.0 & 3.1 \\
CNPC & Fueling & 0.0 & 2.7 \\
Shell & Fueling & 0.0 & 2.3 \\
Car Life & Vehicle insurance, highway toll & 0.0 & 2.8 \\
Halo & Shared bicycle & 12.5 & 2.6 \\
Didi & Taxi & 8.1 & 3.3 \\
Caocao & Taxi & 6.3 & 1.3 \\
Transport & Public transportation & 13.9 & 2.4 \\
Outgoing & Taxi, public transportation & 16.8 & 3.1 \\ \hline
\end{tabular}
\end{adjustbox}
\end{table}

\begin{figure}[!t]
    \centering
    \includegraphics[width=1.\columnwidth]{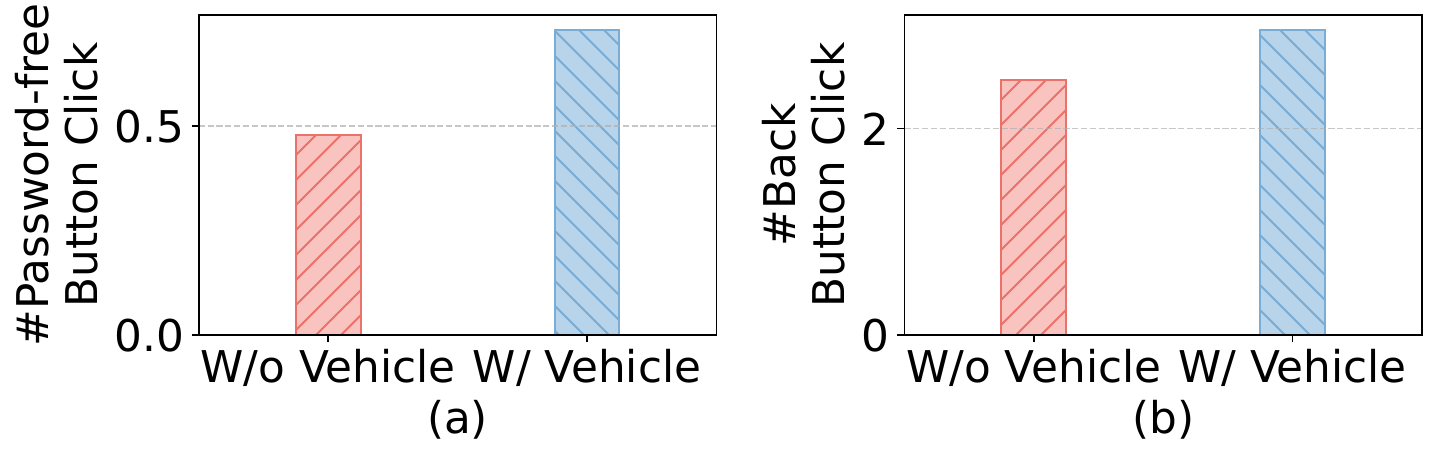}
    \vspace{-2em}
	\caption{Detailed analysis on the privacy of \textit{vehicle}. }
	\label{fig:vehicle}
\end{figure}

\noindent \textbf{Vehicle.} 
Table~\ref{tab:vehicle} shows representative mini-apps that may expose vehicle ownership, the services provided by each mini-app, and the access frequencies of users with and without
vehicles. Users without vehicles rarely access mini-apps for traffic police,
fueling, and vehicle insurance (\eg \texttt{12123}, \texttt{Sinopec},
and \texttt{Car Life}; frequency lower than $0.2$). Contrarily, vehicle users access these mini-apps more frequently (average $2.5$ times). 
Users without vehicles prefer public
transportation mini-apps (\eg \texttt{Transport}, \texttt{Didi} and
\texttt{Caocao}). 
In Table~\ref{tab:vehicle}, \#Access for these users ($11.5$) is significantly higher than vehicle users ($2.5$).
This is because users
without vehicles rely on public transportation or a taxi to travel, while
users with vehicles can drive their cars.

For \op, we show \#Click on two buttons in Figure~\ref{fig:vehicle}: \texttt{Back} and \texttt{Password-free
Payment}. The blue and red bars represent the users with and without vehicles. Users
with vehicles have a high \#Clicks on both buttons, higher than users without
vehicle by an average of 38.55\%.
This may be because users
with vehicles are more likely to use super-apps while
driving~\cite{oviedo2020frustrating}, making password-free payments more
convenient and more frequently using the \texttt{Back} button to return to the
previous page.

\begin{figure}[!t]
    \centering
    \includegraphics[width=1.0\columnwidth]{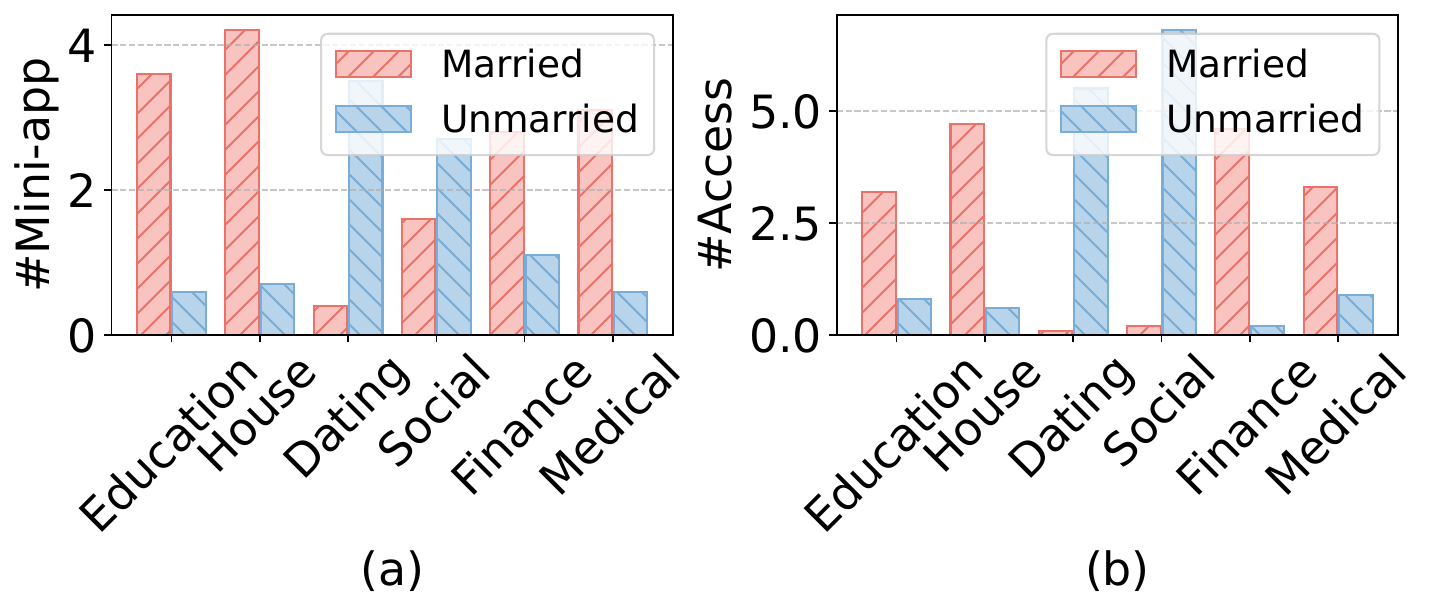}
     \vspace{-2em}
	\caption{\revised{Detailed analysis on the privacy of \textit{marital}.}}
	\label{fig:marital}
\end{figure}

\noindent \textbf{Marital.}
We display \#Mini-app and \#Access for representative mini-app categories in
Figure~\ref{fig:marital}. The married users prefer mini-apps about \texttt{Education}, \texttt{House
and home}, \texttt{Finance}, and
\texttt{Medical}. For these types in Figure~\ref{fig:marital}, the red bars (married users) are significantly higher than
the blue bars (unmarried users) by $4.7\times$ for \#Mini-app and $6.3\times$ for \#Access. This means married users focus more on family
aspects of life. 
Conversely, unmarried users prefer mini-apps about
\texttt{Dating} and \texttt{Social}, with $1.5\times$ higher \#Mini-app and $82.0\times$ higher \#Access than married users.

\begin{figure}[!t]
    \centering
    \includegraphics[width=1.0\columnwidth]{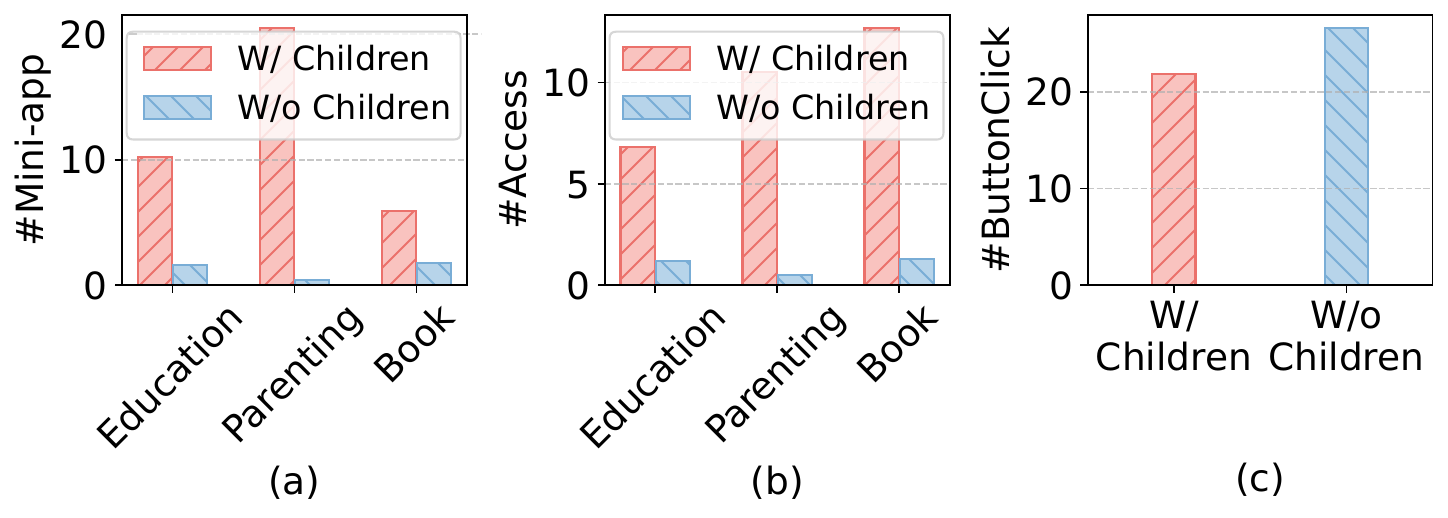}
     \vspace{-2em}
	\caption{Detailed analysis on the privacy of \textit{parental}.}
	\label{fig:parental}
\end{figure}

\noindent \textbf{Parental.} 
We display \#Mini-app and \#Access for representative
mini-app categories in Figure~\ref{fig:parental}~(a) and (b), and show \#ButtonClick in Figure~\ref{fig:parental}~(c).
For \mini, users with children prefer the mini-apps about
\texttt{Education}, \texttt{Parenting}, and
\texttt{Books}. In Figure~\ref{fig:parental}~(a) and
(b), the \#Mini-app and \#Access of users with children (red bars) are
higher than those without children (blue bars) by $9.6\times$ and $10.0\times$,
respectively. We believe the reason is that parents tend to pay more attention
to their children's education.
For \op, users with children have a lower \#ButtonClick than users
without children. In Figure~\ref{fig:parental}~(c), the \#ButtonClick of
parents (red bar) is $18.0\%$ lower than childless users (blue bar).

\subsection{Architecture and Calibration Comparison}

\label{sec:architecture_comparison}

In this section, we study different choices of model architecture and confidence
calibration techniques. For the model architecture, we select three
representative architectures: a CNN-based model, an RNN-based model, and our
Transformer-based model. The CNN-based model is a ResNet model with 48
convolutional layers~\cite{lecun1995convolutional, he2016deep}. The RNN-based
model has 8 LSTM layers following the default architecture of prior
literature~\cite{salehinejad2017recent, hochreiter1997long}. All
architectures have been demonstrated to be effective to extract high-dimensional
information from time series
data~\cite{zerveas2021transformer,hong2020holmes,ma2021effectiveness,ding2019modeling}.
For calibration, we compare the temperature calibration with two
representative techniques: vector calibration and matrix
calibration~\cite{pmlr-v70-guo17a}.

\noindent \textbf{Evaluation Metrics.}
We use four different metrics to evaluate the performance: the proportion of high-confidence data (PHC), precision,
recall, and F1-score. 
Given a confidence threshold, the PHC measures the proportion of data samples that
the model can confidently infer. A higher PHC indicates a higher attack performance.
Precision, recall, and F1-score are three widely used metrics to evaluate classification models. We use these three metrics to
comprehensively evaluate the model's correctness on the PHC data samples.

\noindent \textbf{Results.}
We display the averaged results in Table~\ref{tab:results}, in which the values are
averaged across the seven labels of privacy attributes.
First, for all settings, the PHC is above $8.8\%$. Precision, recall, and F1-score
are all above $89.6\%$. The
proposed attack is general to different architectures
and calibration techniques. 
Second, among all the settings, the Transformer-based model with temperature
calibration achieves the best performance. The PHC is $16.1\%$, and the
precision, recall, and F1-score are all above $95.4\%$. This means that this setting can
correctly infer the privacy of the most number of users with the highest
accuracy.

\newcolumntype{Y}{>{\centering\arraybackslash}X}
\begin{table*}[]
\centering
\caption{Comparison between model architectures and calibration techniques. For each metric, we report the average value across seven privacy attributes.}
\label{tab:results}
\begin{adjustbox}{max width = 0.9\linewidth}
\begin{tabularx}{\linewidth}{Y|YYY|YYY|YYY}
\hline
 Model & \multicolumn{3}{c|}{CNN-based}  & \multicolumn{3}{c|}{RNN-based} & \multicolumn{3}{c}{Transformer-based} \\ \hline
 Calibration & Temperature & Vector & Matrix & Temperature & Vector & Matrix & Temperature & Vector & Matrix \\ \hline
 PHC & 14.6\%  & 11.8\% & 15.9\% & 10.0\% & 7.9\% & 9.8\%  & \textbf{16.1\%} & 15.6\% & 8.8\% \\
 Precision & 91.9\% & 91.3\% & 90.4\% & 91.1\% & 92.2\% & 91.4\% & \textbf{95.5\%} & 94.4\% & 95.3\% \\
 Recall & 91.8\% & 90.6\% & 89.8\% & 90.2\% & 90.5\% & 89.6\% & \textbf{95.4\%} & 94.1\% & 94.4\% \\
 F1-score & 91.8\% & 90.8\% & 89.9\% & 90.5\% & 91.1\% & 90.1\% & \textbf{95.4\%} & 94.2\% & 94.8\% \\ \hline
\end{tabularx}
\end{adjustbox}
\end{table*}

\subsection{Ablation Study}
\label{sec:ablation_study}
In this section, we compare the results of different settings used in \tool.
\begin{figure}[t]
    \centering
    \includegraphics[width=1.0\columnwidth]{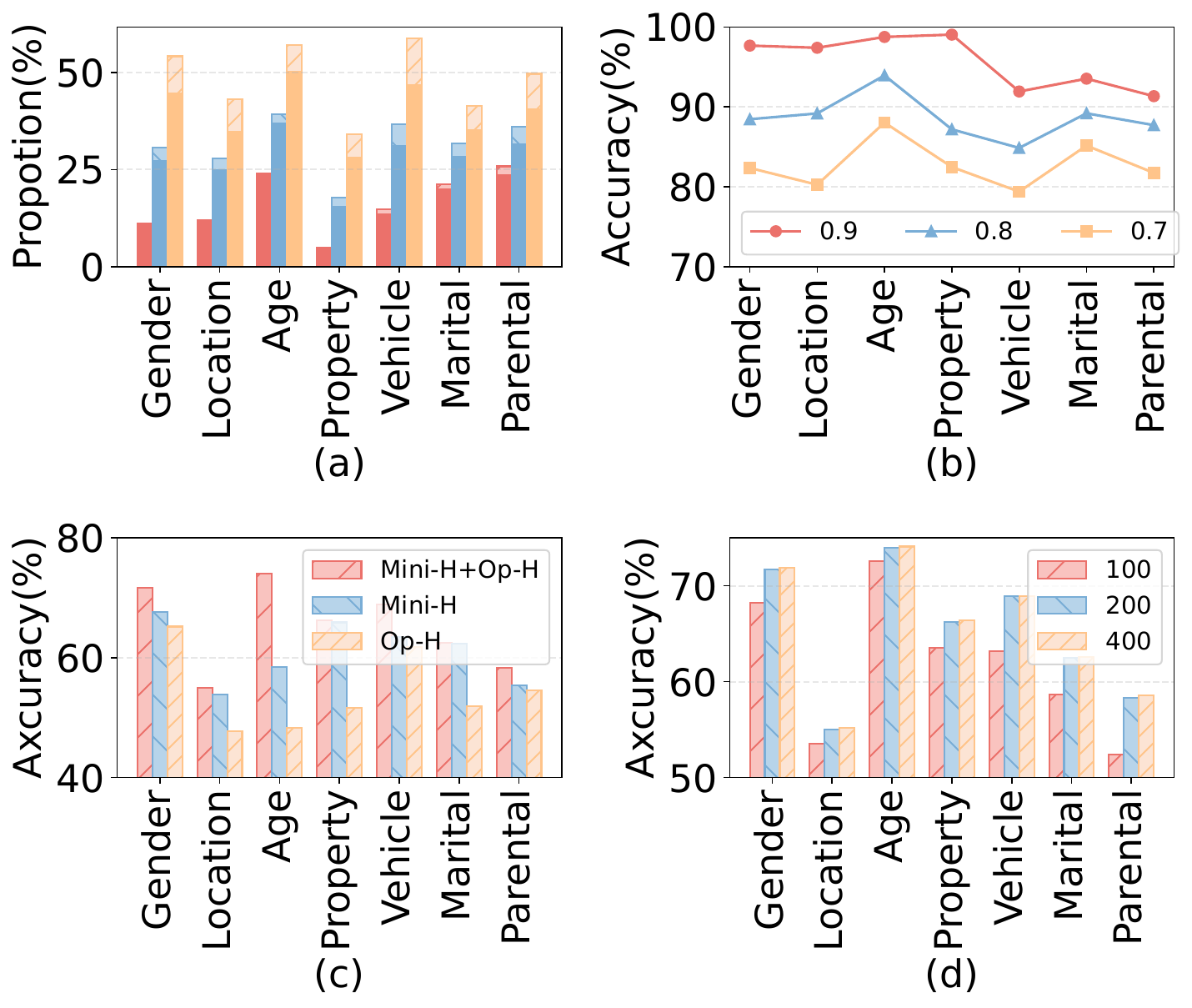}
    \vspace{-2em}
	\caption{Comparision of different settings.}
	\label{fig:appendixsettings}
 \vspace{-1em}
\end{figure}

\noindent \textbf{Threshold.}
First, we assessed the impact of the threshold settings. A higher threshold means we only attack samples with higher confidence scores, implying a smaller victim subset and higher accuracy. Specifically, we compared the proportion of identified samples and their accuracy when setting the threshold at 0.9, 0.8, and 0.7. In Figure~\ref{fig:appendixsettings} (a), light-colored bars represent the proportion of samples that can be identified, and dark-colored bars represent the proportion of accurately inferred samples. We can observe that a lower threshold encompasses more samples, but the accuracy is relatively decreased. The results for the three threshold settings are displayed in Figure~\ref{fig:appendixsettings} (b), where a 0.9 threshold achieves an average accuracy of 95.7\%, while 0.8 and 0.7 only reach average accuracies of 88.7\% and 82.8\% respectively. \revised{These findings demonstrate that higher confidence scores align with higher inference accuracy, thereby validating our model calibration technique. }In this paper, we therefore adopt a threshold of 0.9 to maximize our attack performance.

\noindent \textbf{Input Data.}
\tool's input data is a concatenation of \mini and \op. To further demonstrate that both data types are privacy-related, we trained models using \mini and \op separately and reported the results in Figure~\ref{fig:appendixsettings} (c). Only using \mini or \op achieves an accuracy of 61.1\% and 54.4\%, respectively. Although the values are lower than \tool's concatenation solution (65.2\%), they are higher than the random guess baseline by an average of 44.0\%. Thus, both \mini and \op are effective for \tool.

\noindent \textbf{Data Length.}
In \tool, we empirically set $N=200$. In this section, we compare it with $N=100$ and $N=400$. As shown in Figure~\ref{fig:appendixsettings} (d), using 
$N=400$ only provides a marginal improvement of 0.2\%, while $N=100$ decreases accuracy by 5.4\%. Therefore, our setting is an effective choice.

\section{Industry Feedback}

To help improve the super-app ecosystem, we notified our findings super-app developers and the
privacy standards expert teams of the IEEE Standards
Association (IEEE-SA) in China. We summarize their feedback and critical lessons as follows.

\subsection{Notification Process}

\textbf{Methodology.}
Following prior literature, we contacted super-app developers via email addresses extracted from Apple's App Store submissions and super-apps' official websites. In our emails and
online communications, we briefly explained potential leakage risks and
presented our results. We also posed three crucial questions to
the developers and experts: 
(1) whether they were aware of the privacy implications of the data collected under existing legal regulations,
(2) whether they were aware that such information could infer privacy,
and (3) whether they had any remedial plans or suggestions for addressing these privacy risks.

\noindent \textbf{Results.}
We contacted the vendors of all 31 super-apps in Table~\ref{tab:superapp_summary}
and the standards association. We conducted two rounds of
communication. The first round was completed by October 31, 2023.
A second round was initiated for those who did not reply in the first round and was completed by March 31, 2024. Among the 31 vendors, eight acknowledged our report, and four acknowledged and provided feedback on our
questions. 
Most of the unresponsive vendors either didn't reply or
stated that they would forward the report to relevant teams, after which no
further response was received. We speculate that the lack of response from some
vendors could be due to competitive concerns and the protection of business
secrets.
We also engaged the privacy standards department, and they acknowledged our report. Notably, during our communication with company developers, they showed significant interest in our study and frequently asked for details about our attack.

\subsection{Feedback}

Four super-app vendors and the standards association have committed to updating their privacy policies. This update aims to alert users about the recorded mini-app interaction history data during usage and the potential privacy connections. Moreover, we received the following feedback from the developers and experts:

\noindent \textbf{Feedback 1}: The developers acknowledged that ``\textit{We will fix the notification of potential risks of the mini-app interaction history data in our user terms.}'' They further stated that ``\textit{The identified privacy risks indeed exist. Previously, there was a subconscious belief that collecting mini-app interaction history data does not 
violate regulations due to the absence of studies on the privacy leaks they might cause.}''

\noindent \textbf{Feedback 2}: The developers stated: ``\textit{We are working in progress to protect the mini-app interaction history as other sensitive data.}'' Furthermore, they admitted that they had overlooked the privacy issue of mini-app interaction history by saying: ``\textit{We usually follow the latest research papers to protect users' privacy. However, to our knowledge, there is no paper revealing the privacy risks of this data. Thus, we were not aware of the privacy issue.}''

\noindent \textbf{Feedback 3}: The developers promised to ``\textit{improve protection of mini-app interaction history by reducing the time to store such data.}'' However, developers also mentioned the trade-offs between cost and privacy protection: ``\textit{The computation budget to protect privacy is limited. There is a balance between privacy protection and delivering satisfactory services. The user data generated by super-apps is vast, and the cost to protect all data is immeasurable}.''

\noindent \textbf{Feedback 4}: The developers agreed to improve their user term in collecting mini-app interaction history by acknowledging: ``\textit{Before this study, we considered mini-app interaction history as privacy insensitive because we only simply reviewed the data but didn't conduct in-depth research on it.}''

\noindent \textbf{Feedback 5}: The experts from the standards association noted: ``\textit{During our communications with super-app companies, their developers agreed to add mini-app interaction history into the user terms. The association will also standardize the collection and use of such data in the future.}''

\section{Discussions}

\noindent \textbf{Lessons Learned. }
We identified several key lessons to enhance privacy in super-apps. First, developers must rectify all existing misconceptions regarding the boundaries of privacy security. This includes thoroughly re-examining their services to identify and safeguard potentially vulnerable mini-app interaction history. Second, developers must ensure user data privacy rather than merely reacting to privacy breaches. From a regulatory perspective, standards associations should strive to preemptively address new security vulnerabilities instead of waiting to act after breaches occur.

\noindent\revised{\textbf{Potential Interaction Changes.} One potential concern is that the user agreement could introduce bias, influencing volunteers' interaction patterns with the mini-apps. However, we believe this bias has a negligible impact due to the large volume of data, which helps mitigate individual biases. Nevertheless, we acknowledge that measuring bias is challenging since we cannot access ground-truth data (e.g., usage history outside the experimental period or from non-participants).}

\noindent \textbf{Performance of Attacks.}
Note that even though our focus is not on all users, the implications of accurate predictions on a high-confidence subset are profound. This subset, albeit a fraction of the overall users, includes many people due to the colossal scale of super-app users. For instance, for a super-app with over a billion users, accurately inferring privacy attributes of even 1\% users will impact 10 million individuals. This magnitude is substantial and raises serious concerns on personal privacy. Our research highlights this issue and draws attention to robust privacy protection measures, even when only a subset of users is identified.

\noindent \textbf{Privacy Statement Revision.} In the feedback, five super-app vendors promised to revise their user terms. This marks a milestone in super-app privacy, as these industry giants acknowledge and address the privacy implications of mini-app interaction history. By updating the privacy statements, these vendors have demonstrated a commitment to enhancing user privacy and set a precedent for others. The willingness of these vendors to adapt and improve their privacy statements in light of our research is also a strong testament to the generalizability of our findings.

\noindent \textbf{Generalizability.} A key aspect of our research is its broad generalizability, not just to \alipay but to various super-apps as well. The unified interface design and operational paradigms of super-apps and mini-apps lead to the generality of \mini and \op. Therefore, the privacy vulnerabilities we identified from the mini-app interaction history are generalizable findings.

\noindent \textbf{Defense Techniques.} Common defense methods include differential privacy or trusted hardware. However, applying differential privacy to mini-app interaction history could lead to information loss and mislead user interactions with other operations. Utilizing trusted hardware like SGX~\cite{273705} significantly increases costs, which is unsustainable for super-apps with hundreds of millions of active users. Therefore, there is a need to discover more effective defense methods.

\section{Related Work}
Previous research has witnessed a significant focus on privacy protection in the mobile area. 
These works highlight the intricate privacy dimensions specific to mobile apps. Specifically, these studies offer valuable perspectives on several key areas: they explore the privacy challenges in super-app ecosystems~\cite{dongexploring, 10.1145/3576915.3616676, 10.1145/3576915.3616591, 10.5555/3620237.3620608, yang2022cross, zhang2022identity, haney2021s, chen2019demystifying}, investigate methods through which data might be stealthily exfiltrated or mishandled~\cite{nan2023are, koch2023ok,diamantaris2021sneaky}, assess the alignment between the mobile app and regulations~\cite{pantrap, khandelwal2023unpacking, 10.1145/3576915.3623067, xiao2023lalaine, meng2023post, li2022collect, yang2022cross, young2022skilldetective, balash2022security, bui2021consistency, nguyen2021share, zuo2019does}, and propose strategies to enhance privacy compliance and transparency within apps~\cite{klein2023general,ferreira2023rulekeeper,jordan2021viceroy,nguyen2022freely,wang2022srr,cai2024famos}.
These studies emphasize the complexity of privacy issues, covering personal and device information, location, camera, microphone, etc. These investigations reflect the community's commitment to tackling mobile app privacy and security challenges. %

However, an unexplored area remains regarding the potential privacy risks in the mini-app interaction history within super-app ecosystems. Unlike location or microphone data, these data do not typically trigger system-level permission prompts, making them less visible and hence less concerned. Meanwhile, they can reveal extensive insights into user privacy attributes. In this paper, we close a critical gap in understanding how seemingly non-sensitive mini-app interaction history can pose novel risks, underscoring the need for more nuanced privacy protections for the super-app ecosystems.

\section{Conclusion}

This paper reveals a new super-app privacy vulnerability: the mini-app interaction history. We design a new attack, \tool, that can achieve more than 95.5\% accuracy in inferring privacy attributes of {over 16.1\% of users} with less than 0.1\% training data. We also highlight a significant oversight in the academic community and industry practitioners on protecting mini-app interaction history. Our findings have also raised awareness and proactive measurements among super-app vendors and standards associations.

\section*{Acknowledgments}
We would like to thank the anonymous reviewers for
their valuable feedback. 
Ding Li is the corresponding author. 
This work was partly supported by the National Science and Technology Major Project of China (2022ZD0119103) and the CCF-Ant Research Fund (RF20220006).

\section*{Ethics Considerations}
\label{sec:ethics_star}
\revised{We carefully reviewed the conference’s ethical guidelines, submission instructions, and ethics documents. Our research was conducted ethically and responsibly, with IRB approval from \alipay obtained before starting the study.}

\revised{We conducted our study transparently, assessing and mitigating risks for all stakeholders. For volunteer participants, we secured informed consent, anonymized their data, and ensured secure handling to prevent unauthorized access. A dedicated team of internal engineers, under strict confidentiality agreements, managed data extraction and anonymization, providing researchers only with anonymized, non-identifiable datasets.}

\revised{For \alipay as a company and its employees, there was a potential risk of reputational damage or exploitation of identified vulnerabilities by insiders. We proactively engaged with relevant departments to address these vulnerabilities and provided recommendations for improved security practices, aiming to minimize any negative impact on the company.}

\revised{Considering the broader super-app user base, we recognized that our findings might reveal vulnerabilities that could be exploited if not properly addressed. To mitigate this risk, we disclosed our findings to \alipay's security team and other major super-app vendors. We received positive responses, and several vendors have committed to addressing these issues, thereby enhancing the security of all users.}

\noindent \revised{\textbf{\alipay's IRB Overview.} \alipay, an international company with over a billion users, has an IRB that complies with ethical standards like the Common Rule and ISO 27701 for privacy management. This independent committee includes legal experts, ethicists, and external professionals unaffiliated with the company, ensuring unbiased and thorough reviews. The IRB evaluates all research involving human subjects to uphold the highest standards of informed consent, confidentiality, and data protection.}

\noindent \revised{\textbf{Participant Data Protection within \alipay. }We protected participant data both externally and within \alipay by restricting access to a dedicated team of internal engineers bound by strict confidentiality agreements. This team handled data extraction and anonymization, with all activities logged and monitored. Engineers were prohibited from using the data for any purposes beyond this study. Researchers received only anonymized datasets with pseudonyms replacing unique identifiers, ensuring no access to personally identifiable information. Additionally, the engineers did not participate in research analysis, maintaining a clear separation of roles and further safeguarding participant privacy.}

\noindent \revised{\textbf{User Consent and Data Collection.} Since our evaluation involves the collection of privacy attributes, we obtained IRB approval from \alipay before collecting any data. The IRB requires us to obtain volunteers' consent prior to data collection, and all procedures were conducted by internal engineers at \alipay under IRB supervision.}

\revised{For volunteer selection, internal engineers sent invitations to active users using automated systems that only checked login timestamps, ensuring no sensitive information was accessed. This procedure was approved by the IRB as low-risk and compliant with data protection policies. Once volunteers were selected, the engineers assisted us with the following steps:}

\revised{$\bullet$ Informed Consent: We informed the volunteers about the data we would collect, its intended use, and the measures taken to protect it. We assured them that the data would not be used for commercial purposes or shared with third parties.}

\revised{$\bullet$ Consent Acquisition: We obtained consent from users to access their mini-app interaction history and privacy attributes. For adult users, we obtained their consent directly. For minors, who typically use authorized accounts, we obtained consent from both the users and their parents.}

\revised{$\bullet$ Attributes Verification: During data verification, internal engineers matched participants' provided attributes with internal databases(e.g., databases from the loan and insurance departments) in a secure and controlled environment. The matching process was automated to minimize human exposure to sensitive data.}

\revised{$\bullet$ Data Collection: We used \alipay-dev to notify volunteers of their participation, after which internal engineers extracted their mini-app interaction history from \alipay's servers}

\revised{Note that the internal engineers anonymized all unique user identifiers before being provided to us. The data was securely stored in an encrypted database and was carefully deleted under IRB supervision after the evaluation.}

\noindent \revised{\textbf{Data Access.} All code and anonymized data were accessible only to the authors. The NDA approval process involved submitting a data access request to \alipay's IRB, which reviewed the request for compliance with data protection policies. An agreement was then signed by the authors and authorized personnel, and upon NDA approval, data access permissions were granted. Overall, all data access and handling complied with \alipay's data protection policies and relevant laws, ensuring participant privacy and data security throughout the research.}

\section*{Open Science}
\revised{
We recognize the importance of open science and are committed to supporting the research community. However, due to \alipay's IRB and business regulations, and strict user privacy concerns, all code and data are processed on supervised servers. Therefore, we cannot publicly release certain artifacts, including code and comprehensive datasets. Meanwhile, we have provided detailed descriptions of our work, including design, methods, input formats, configurations, etc. In addition, we are committed to helping colleagues who wish to replicate or fully understand our work. We encourage contacting us to discuss potential collaborations or access materials under an NDA and with approval from \alipay.}
\bibliographystyle{plain}
\bibliography{bibs}

\end{document}